\documentclass[journal]{IEEEtran}
\IEEEoverridecommandlockouts

\usepackage{amsmath,amssymb,amsfonts,amsthm,bm,mathrsfs}

\usepackage{graphicx}
\usepackage[dvipsnames]{xcolor}
\usepackage{tikz}
\usetikzlibrary{shapes.geometric, positioning, backgrounds}
\usepackage[export]{adjustbox}

\usepackage{booktabs}
\usepackage{multirow}
\usepackage{textcomp}
\usepackage{comment}
\usepackage{caption}
\usepackage{subcaption}

\usepackage{algorithmic}
\usepackage[ruled,vlined]{algorithm2e}
\usepackage{array}

\usepackage{cite}
\usepackage[acronym]{glossaries}
\usepackage{xifthen}
\usepackage{etoolbox}
\usepackage{nomencl}

\newtheorem{definition}{Definition}

\renewcommand\nomgroup[1]{%
  \item[\bfseries
  \ifstrequal{#1}{P}{Parameters}{%
  \ifstrequal{#1}{S}{Sets}{%
  \ifstrequal{#1}{V}{Variables}{}}}%
]}

\newcommand{\tikzxmark}{%
\tikz[scale=0.23] {
    \draw[line width=0.7,line cap=round] (0,0) to [bend left=6] (1,1);
    \draw[line width=0.7,line cap=round] (0.2,0.95) to [bend right=3] (0.8,0.05);
}}
\newcommand{\tikzcmark}{%
\tikz[scale=0.23] {
    \draw[line width=0.7,line cap=round] (0.25,0) to [bend left=10] (1,1);
    \draw[line width=0.8,line cap=round] (0,0.35) to [bend right=1] (0.23,0);
}}

\makenomenclature

\newcommand{\Sajjad}[1]{{\color{Black} #1}}

\begin{document}

\title{Distribution System Reconfiguration to Mitigate Load Altering Attacks via Stackelberg Games}
\author{\IEEEauthorblockN{Sajjad Maleki, \textit{Graduate Student Member, IEEE}, E. Veronica Belmega, \textit{Senior Member, IEEE}, Charalambos Konstantinou, \textit{Senior Member, IEEE}, and Subhash Lakshminarayana, \textit{Senior Member, IEEE}
}
\vspace{-1em}


\thanks{S. Maleki is with the School of Engineering, University of Warwick, CV47AL, UK and ETIS UMR 8051, CY Cergy Paris Universit\'e, ENSEA, CNRS, F-95000, Cergy, France. E. V. Belmega is with Univ. Gustave Eiffel, CNRS, LIGM, F-77454, Marne-la-Vallée, France and ETIS UMR 8051, CY Cergy Paris Universit\'e, ENSEA, CNRS, F-95000, Cergy, France. C. Konstantinou is with the CEMSE Division, King Abdullah University of Science and Technology (KAUST).  S. Lakshimnarayana is with the School of Engineering, University of Warwick, CV47AL, UK. Emails: (sajjad.maleki@warwick.ac.uk, veronica.belmega@esiee.fr, charalambos.konstantinou@kaust.edu.sa, subhash.lakshminarayana@warwick.ac.uk).

This work has been supported in part by the PhD Cofund WALL-EE project between the University of Warwick, UK and CY Cergy Paris University, France and in part by the King Abdullah University of Science and Technology (KAUST) under Award No. RFS-OFP2023-5505.
{This work was partially presented at IEEE PES General Meeting-2024 \cite{maleki2023impact}.}}
}

\maketitle

\nomenclature[P]{\(\alpha_q , \beta_q , \gamma_q\)}{Reactive ZIP load coefficients}
\nomenclature[P]{\(\alpha_p , \beta_p , \gamma_p\)}{Active ZIP load coefficients}
\nomenclature[P]{\(\alpha'_p, \gamma'_p\)}{Active ZP load coefficients}
\nomenclature[P]{\(\alpha'_q , \gamma'_q\)}{Reactive ZP load coefficients}
\nomenclature[V]{\(p_i^0\), \(p_i^0\)}{Active and reactive load demands at rated voltage in bus $i$}
\nomenclature[V]{\(v_i\)}{Voltage in bus $i$}
\nomenclature[P]{\(v_{nom}\)}{Nominal voltage}
\nomenclature[P]{\(r_{ij}, x_{ij}\)}{Resistance and reactance of the line from bus $i$ to $j$}
\nomenclature[V]{\(b_{ij}\)}{Component of $B$ in $i^{th}$ row and $j^{th}$ column}
\nomenclature[S]{\(\mathcal{N}^f\)}{Subset of buses which are substation}
\nomenclature[S]{\(\mathcal{N}\setminus \mathcal{N}^f\)}{Subset of buses which are not substations}
\nomenclature[S]{\(\mathcal{L}^s\)}{Set of lines with switches}
\nomenclature[S]{\(\mathcal{L}\setminus \mathcal{L}^s\)}{Set of lines without switches}
\nomenclature[S]{\(\mathcal{N}^a\)}{Subset of buses under attack}
\nomenclature[V]{\(p_{ij} , q_{ij}\)}{Active and reactive power flows from bus $i$ to $j$}
\nomenclature[V]{\(p_i^f, q_i^f\)}{Active and reactive powers flowing from substation bus $i$}
\nomenclature[P]{\(M\)}{Disjunctive parameter}
\nomenclature[V]{\(\hat{v}_{ji}\)}{Auxiliary voltage variable for MILP}
\nomenclature[V]{\(p^a, q^a\)}{Active and reactive powers raised by LAA}
\nomenclature[V]{\(a^*, r(a^*)\)}{Players actions in equilibrium}
\nomenclature[V]{\(r(a)\)}{Follower's best response to attacker's strategy}
\nomenclature[V]{\(n_{att}\)}{The bus under attack}
\nomenclature[V]{\(s_i^{\text{zip}}, s_i^{\text{zp}}\)}{Apparent power in bus $i$ with ZIP and ZP load models}
\nomenclature[V]{\(s_i\)}{Apparent power in bus $i$}
\nomenclature[V]{\(B\)}{Adjacency matrix of the distribution network after reconfiguration}
\nomenclature[P]{\(b_{ij}^{pre}\)}{Component of the adjacency matrix before reconfiguration in $i^{th}$ row and $j^{th}$ column}
\nomenclature[V]{\(\pi_i\)}{Parent bus of the bus $i$}
\nomenclature[V]{\(\sigma_{\ell}\)}{Probability of the bus $\ell$ to be under attack}
\nomenclature[S]{\(\mathcal{D}_i\)}{Set of buses engaged in the unique path connecting the bus $i$ to the root bus}
\nomenclature[P]{\(p_i^l, q_i^l\)}{Active and reactive load demands in bus $i$}
\nomenclature[P]{\(p_d, q_d\)}{Nominal active and reactive power demands by a single attacked device}
\nomenclature[V]{\(\textbf{U}\)}{Vector of the square of voltages of the system while no LAA}
\nomenclature[V]{\(\textbf{U}^A\)}{Vector of square of voltages of the system under LAA}
\nomenclature[P]{\(N\)}{Number of buses of the distribution system}
\nomenclature[P]{\(\textbf{I}\)}{Identity matrix}
\nomenclature[V]{\(\Omega , \Omega' , \Omega''\)}{Coefficients matrices for the proposed closed-form expressions}
\nomenclature[V]{\(c(n)\)}{Number of attacked devices for a ``critical attack" in bus $n$}
\nomenclature[V]{\(I_{\pi_i, i}\)}{Current flowing through the branch $(\pi_i,i) \in \mathcal{L}$}
\nomenclature[V]{\(p_{\text{der},i}, q_{\text{der},i}\)}{Active and reactive power output of DER in bus i}
\nomenclature[P]{\(\overline{s_{\text{der},i}}\)}{Maximum output capacity of DER in bus i}
\nomenclature[S]{\(\mathcal{N}^{\text{DER}}\)}{Set of buses with DER}
\begin{abstract}

\Sajjad{The widespread integration of IoT-controllable devices (e.g., smart EV charging stations and heat pumps) into modern power systems enhances capabilities but introduces critical cybersecurity risks. Specifically, these devices are susceptible to load-altering attacks (LAAs) that can compromise power system safety. This paper quantifies the impact of LAAs on nodal voltage constraint violations in distribution networks (DNs). We first present closed-form expressions to analytically characterize LAA effects and quantify the minimum number of compromised devices for a successful LAA. Based on these insights, we propose a reactive defense mechanism that mitigates LAAs through DN reconfiguration. To address strategic adversaries, we then formulate defense strategies using a non-cooperative sequential game, which models the knowledgeable and strategic attacker, accounting for the worst-case scenario and enabling the reactive defender to devise an efficient and robust defense. Further, our formulation also accounts for uncertainties in attack localization. A novel Bayesian optimization approach is introduced to compute the Stackelberg equilibrium, significantly reducing computational burden efficiently. The game-theoretic strategy effectively mitigates the attack's impact while ensuring minimal system reconfiguration.}
\end{abstract}
 \begin{IEEEkeywords}
 Distribution network, Cybersecurity, Load-altering attack (LAA), Reconfiguration, Stackelberg game, Bayesian optimization.
 \end{IEEEkeywords}
\Sajjad{\printnomenclature[4em]}
\section{Introduction}

Internet-of-Things (IoT) devices offer enhanced end-user convenience, improved efficiency and flexibility to power systems for load peak management, which has driven a notable surge in their adoption. However, beyond their evident benefits, these devices also present potential vulnerabilities, serving as entry points for cyber attackers to compromise the security of power systems. Specifically, load-altering attacks (LAA) in power networks with high IoT-enabled device penetration pose a significant cybersecurity threat \cite{mohsenian2011distributed, amini2016dynamic, soltan2018blackiot, lakshminarayana2021analysis}. 

\Sajjad{The concept of LAAs was first introduced in \cite{mohsenian2011distributed}. In this work, adversaries turn a group of IoT-controllable electrical loads into bots, which are then turned on and off simultaneously to harm the stability of the system.} The manipulation of loads disrupts the balance between power generation and demand, leading to frequency instabilities in transmission networks \cite{dabrowski2017grid, soltan2018blackiot}. In distribution networks (DNs), LAA can result in elevated line flows, causing higher voltage drops, leading to voltage constraint violation \cite{liu2021robust}. Furthermore, the data required to launch \Sajjad{a} successful LAA can be obtained from publicly available information \cite{acharya2020public} or \Sajjad{bypassed} \cite{soleymani2024data}.

\subsection{Literature Survey} \label{lit_survey} 
LAAs have gained significant interest over the last few years.  
We divide the existing works into two groups: 1) attack impact analysis and viability; and 2) attack mitigation.

\textbf{Attack Impact Analysis and Viability:}
 Researchers in \cite{soltan2018blackiot} investigated the impact of LAA on transmission systems and identified several effects, including line failure, unsafe frequency deviation, disruption in grid restarting, and increased operational cost.  
 \Sajjad{The research in \cite{huang2019not} examined the effects of LAA under a more realistic setting. This setting included protection schemes (such as $N-1$ security) and load-shedding schemes. The study showed that even in this context, LAA can still cause outages and islands. Additionally, a recent work \cite{rodriguez2025madiot} expands the LAA model and launches the attack on high-wattage IoT devices and distributed energy resources (DERs) simultaneously. One notable result from their work is that attacking scattered devices causes a lower impact than attacks on devices in a concentrated area.}

Reference \cite{amini2016dynamic} proposed the dynamic LAA (DLAA), in which the adversary continuously toggles the compromised load devices on and off, guided by a feedback control loop in response to the system's frequency fluctuations. 
In \cite{lakshminarayana2021analysis}, an analytical framework was introduced to analyze the impact of LAA on transmission systems and identify the nodes from which an attacker can launch the most effective attacks using the theory of second-order dynamical systems. In \cite{goodridge2024uncovering}, a rare-event sampling algorithm was proposed that uncovers the spatial and temporal distribution of impactful DLAAs while considering the security constraint of $N-1$.

The growing popularity of electric vehicles (EVs) and the spread of EV charging stations (EVCS) make them a potential target for the LAA. Reference \cite{abazari2024electric} \Sajjad{explored} the weak points of EVCS, such as firmware flaws, vulnerabilities in management systems, and the security scarcity of mobile apps that allow adversaries to launch LAAs. Subsequently, to detect such attacks, they propose model-based and data-driven approaches.

\textbf{Attack Mitigation:} Another stream of research investigates the mitigation of LAAs. The existing mitigation methods can be categorized into: i) offline, or ii) online methods.

\begin{table}[t]
  \caption{\small State of the art on LAA mitigation methods and position of our work.}
    \centering
    \begin{tabular}{|>{\centering\arraybackslash}p{0.4cm}|>{\centering\arraybackslash}p{0.5cm}|>{\centering\arraybackslash}p{1cm}|>{\centering\arraybackslash}p{1cm}|>{\centering\arraybackslash}p{2cm}|>{\centering\arraybackslash}p{1.3cm}|} \hline
        &\multirow{2}{*}{Ref} & {Impact analyses} & Strategic attacker & \multirow{2}{*}{Defense approach} & Defense type \\ \hline
        \multirow{13}{*}{\rotatebox[origin = c]{90}{Transmission}}&\cite{amini2016dynamic} & \tikzxmark & \tikzxmark & Securing loads  & Preventive  \\ \cline{2-6}
        &\multirow{2}{*}{\cite{soltan2019protecting}} & \multirow{2}{*}{\tikzxmark} & \multirow{2}{*}{\tikzxmark} & Robust operating points & \multirow{2}{*}{Preventive}  \\ \cline{2-6}
        &\multirow{2}{*}{\cite{chu2022mitigating}} & \multirow{2}{*}{\tikzxmark } & \multirow{2}{*}{\tikzxmark} & Frequency droop control & \multirow{2}{*}{Reactive}  \\ \cline{2-6}
         &\multirow{2}{*}{\cite{sayed2023protecting}} & \multirow{2}{*}{\tikzxmark} & \multirow{2}{*}{\tikzxmark} & EV charge/discharge  & \multirow{2}{*}{Reactive}  \\ \cline{2-6}
          & \multirow{1}{*}{\cite{guo2021reinforcement}} & {\tikzxmark} & {\tikzcmark} & Load shedding & Reactive \\ \cline{2-6}
           &\multirow{3}{*}{\cite{zhao2024integrated}} & \multirow{3}{*}{\tikzxmark} & \multirow{3}{*}{\tikzcmark} & Securing devices \& Optimal load management  & \multirow{3}{*}{Hybrid}  \\ \cline{2-6}
        &\multirow{2}{*}{\cite{an2023robust}} & \multirow{2}{*}{\tikzxmark} & \multirow{2}{*}{\tikzcmark} & Reactive power compensation  & \multirow{2}{*}{Preventive} \\ \hline
         \multirow{3}{*}{\rotatebox[origin = c]{90}{Dist.}} &\cite{liu2021robust} & \tikzxmark & \tikzxmark & SOPs & Preventive \\ \cline{2-6}
        &Our work  & \multirow{2}{*}{\tikzcmark} & \multirow{2}{*}{\tikzcmark} & \multirow{2}{*}{Reconfiguration} & \multirow{2}{*}{Reactive}\\ \hline
    \end{tabular}
    \vspace{-2em}
    \label{comparison}
\end{table}

\textit{Offline methods}:
Offline defenses try to install \emph{preventive} measures to stop the impact of LAA. For instance, \cite{soltan2019protecting} proposed algorithms to determine the operating points for generators in a way to prevent line overloads caused by potential botnet-type attacks against IoT devices.
Reference \cite{amini2016dynamic} presented a mitigation framework based on securing the most critical loads. This method found the minimum magnitude of loads needed to be protected in order to guarantee frequency stability in the event of DLAAs. \cite{an2023robust} proposed a zero-sum Stackelberg game formulation to install reactive power compensation (RPC) to reduce the impact of the attack.
While the works above focus on transmission systems, \cite{liu2021robust} introduced a mitigation approach tailored specifically for DNs. Their research focuses on identifying optimal locations for deploying soft open points (SOPs) and refining their operation to mitigate the effects of attacks on voltage deviations.

\textit{Online methods}: Despite the effectiveness of the offline methods, these measures may be too costly as the preventive features must be enabled irrespective of whether an attack occurs or not (e.g., uneconomic generator operating points to cover for LAAs). Online methods, on the other hand, involve determining defensive actions to counter the effects of LAAs once the attack is launched, via reactive measures. In \cite{chu2022mitigating}, a cyber-resilient economic dispatch method is introduced to mitigate LAAs based on altering the frequency droop control parameters of inverter-based resources to counter the destabilizing effects of LAAs.
\cite{sayed2023protecting} proposed a framework in which electric vehicles are designed as feedback controllers that can mitigate the impact of LAA based on $H-2$ and $H-\infty$ norms.
To analyze the manoeuvres of a strategic attacker initiating DLAA, \cite{guo2021reinforcement} introduced a multi-stage game approach. In this game, the defensive actions involve load shedding, and the ultimate objective is to achieve a strategic balance between DLAA and the necessary amount of load shedding, reaching a Nash equilibrium (NE).
\cite{zhao2024integrated} formulated the cybersecurity of the power system in both cyber and physical layers as a game-theoretic formulation.

\Sajjad{Existing research on LAAs has largely focused on transmission grids. However, cybersecurity research on DNs extends beyond LAAs. One group of studies investigates attack models and vulnerabilities in DNs, aiming to identify potential threats and prepare for them as the \textit{pre-attack} stage. These include assessing load redistribution attacks in unbalanced DNs \cite{choeum2020vulnerability}, proposing a dynamic false data injection attack \cite{lu2022differential}, and highlighting critical operations and components that are prone to cyber-physical attacks \cite{khalaf2024survey}.

A second group of studies addresses the \textit{during-attack} stage by developing detection and mitigation frameworks. This includes hierarchical methods leveraging waveform measurements \cite{li2022adaptive}, unsupervised adversarial autoencoders for attack detection \cite{zideh2024unsupervised}, and vehicle-to-grid (V2G) voltage control schemes for attack mitigation \cite{an2024robust}.

A third group focuses on the \textit{post-attack} stage. For example, \cite{ali2024securing} proposes a high-level three-step grid restoration framework consisting of (1) post-attack detection and localization, (2) isolation and pre-recovery, and (3) recovery and assessment. Furthermore, \cite{liu2025cyber} studies optimal crew routing during the cyber-recovery phase.

Distinct from these works, this paper first analyzes the impact of LAAs on DNs. Based on the insights gained and through a game-theoretic study, a bespoke defense mechanism is proposed that leverages the existing infrastructure of the grid.}

\Sajjad{\textbf{Differences of Attack on Transmission and Distribution Grids:} Due to the radial structure of DNs, an attacker can cause severe voltage drops by compromising a small number of devices at deep (leaf) nodes. For example, in \cite{huang2019not}, the authors conclude that an effective LAA on the PowerWorld 9-bus transmission grid needs to alter 30\% of its total load, which is equivalent to 24682 MW (the power consumed by millions of air conditioners). However, the analysis in this paper shows that a successful attack on the 33-bus DN can be as small as compromising less than 100 air conditioners. This spatial sensitivity does not exist in the same form in transmission systems. Additionally, this spatial sensitivity enables us to propose a reconfiguration-based \cite{mahdavi2023robust} defense in the game-theoretic analysis, whereas previous transmission-level games focus on load shedding or generator dispatch. These diverse defense models also stem from the different threatened stabilities. In transmission grids, LAA primarily threatens frequency stability, often necessitating frequency-based or generator-side mitigation strategies. However, in DNs, LAA mainly affects voltage regulation.

Furthermore, novel DNs have higher dependence on the information technology section, and this makes them a prime target for attackers \cite{khalaf2024survey}. Transmission grids typically have robust protection infrastructures; however, a single successful breach can have catastrophic consequences, such as widespread blackouts. In contrast, cyberattacks on DNs can be launched more frequently and with greater feasibility, though a single incident is less likely to cause large-scale disruption.}
{\subsection{Research Gap and Contributions}} 

\Sajjad{Despite the growing threat of LAAs, existing research mainly focuses on transmission systems, with limited works addressing their impact or mitigation in DNs. Moreover, few studies account for strategic attackers who adapt based on system defenses. This paper addresses these gaps by analyzing the effects of LAA in DNs and proposing a game-theoretic defense strategy based on network reconfiguration.

This paper significantly extends our preliminary work \cite{maleki2023impact} and addresses these challenges and proposes a defense strategy specifically tailored for DN, highlighting several conceptual innovations:

$\bullet$ Deriving closed-form expressions to assess how LAA affects nodal voltages and to quantify the minimum number of IoT-controlled devices needed to cause voltage violations in DNs with voltage-dependent loads.

$\bullet$ Developing a Stackelberg game to model interactions between a strategic attacker and the DN operator, incorporating uncertainty in attack localization.

$\bullet$ Introducing a Bayesian optimization method to compute the Stackelberg equilibrium efficiently, significantly reducing the computational burden of finding optimal reconfiguration strategies.

These contributions collectively enable a practical and scalable framework for analyzing and mitigating LAA in DNs.
}


{The rest of the paper is structured as follows: Section \ref{Preliminaries} presents the implemented models and definitions in the paper. Section \ref{effects} provides an analytical study on the effects of LAAs on distribution systems. Subsequently, \ref{LAAmitigation} presents the proposed LAA mitigation scheme. Section \ref{results} outlines the obtained numerical results and provides discussions. Finally, Section \ref{conclusion} concludes the paper.}
\begin{figure}[t]
    \begin{center}
    \includegraphics[width=0.35\textwidth]{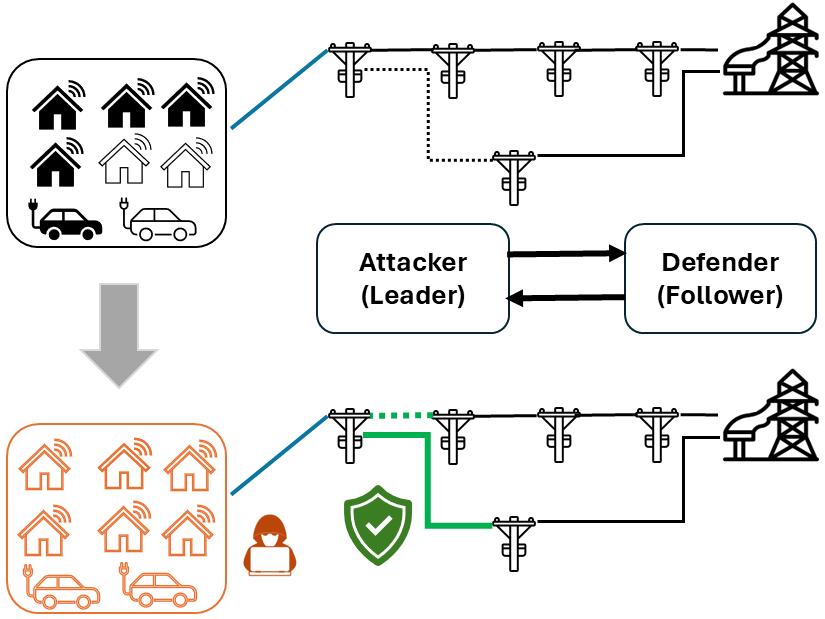}
    \end{center}
    \caption{\small {Summary of the proposed attacker-defender interaction.}}
    \label{flowchart}
    
\end{figure} 
\section{Preliminaries} \label{Preliminaries} 
In this section, we introduce the system load and power flow models for the DNs considered in this research.
\vspace{-1em}
\subsection{DN Model}
The DN under study is represented by the connected directed graph $G = \{\mathcal{N}, \mathcal{L}\}$, where $\mathcal{N} = \{1, 2, \hdots, N\}$ denotes the set of buses and $\mathcal{L}$ denotes the set of branches. This graph has a radial structure, hence it is a tree. Except for bus $1$, which is the root, each bus is referred to as the `child' of its parent bus, which is the adjacent bus closer to bus $1$ by one branch. Thus, the set of branches is defined as $\mathcal{L} = \{(\pi_i, i)\ | \ \pi_i, \ i \in \mathcal{N}\}$. In this configuration, bus $1$ represents the generator bus. 
We denote by $\mathcal{D}_k$ the set of buses which form the unique path connecting bus $1$ to bus $k$, excluding bus $1$ and including bus $k$. The depth of each bus represents the distance in terms of the number of branches between that bus and the root bus.
\subsection{Load Model} 
This subsection introduces the load models which are implemented in the rest of the paper.
\subsubsection{ZIP Load Model}
ZIP and exponential load models are the most implemented ones in the industry. Experimental values for \Sajjad{the} voltage dependency of loads are captured and fitted in the ZIP model in \cite{bokhari2013experimental}, which we integrate into our formulations. The power demand under the ZIP load model is given in \cite{van2007voltage} as follows:
\begin{equation}
     s^{\text{zip}}_i(v_i) = p^l_i (\alpha_p + \beta_p v_i + \gamma_p v^2_i) + j q^l_i (\alpha_q +\beta_q v_i + \gamma_q v^2_i),
     \label{ZIP} 
\end{equation}
where $\alpha_k + \beta_k + \gamma_k = 1$, where $k = \{ p,q\}.$ The ZIP load model captures the voltage dependency of real-world loads.
\subsubsection{ZP Approximation} Based on \eqref{ZIP}, the ZIP model is a function of both {$v_i$ and $v^2_i$}. This causes the optimization tasks involving the power flow in the presence of ZIP loads to become nonconvex and complex. To tackle this problem, \cite{nazir2020approximate} provided an approximate model for ZIP loads given by 
\begin{equation} 
    s^{\text{zp}}_i(v_i) = p^{l}_i (\alpha'_p  + \gamma'_p v_i^2) + j q^{l}_i (\alpha'_q + \gamma'_q v^2_i),
     \label{ZP}
\end{equation}
where
$\alpha'_p = \alpha_p + \frac{\beta_p}{2}$
,$\alpha'_q = \alpha_q + \frac{\beta_q}{2}$
,$\gamma'_p = \gamma_p + \frac{\beta_p}{2}$
, and $\alpha'_q = \gamma_q + \frac{\beta_q}{2}$, while $s^{\text{zp}}_i(v_i) = p_i^{\text{zp}}(v_i) +jq_i^{\text{zp}}(v_i)$. The new coefficients in the ZP model are obtained by the binomial approximation method. The ZP approximation is valid as long as the voltage is close enough to the nominal value, i.e., while $|v_i - v_{nom}| \leq 0.1$, the ZP approximation is valid \cite{nazir2020approximate} ($v_{nom}$ is the nominal voltage). \Sajjad{To further clarify this, we illustrate the dependence of apparent power on voltage using the ZIP and ZP models for an air conditioner in Figure \ref{zipvszp}. As shown, when the voltage is close to 1 p.u., the ZP model produces values very similar to those of the ZIP model. According to \cite{nazir2020approximate}, the ZP approximation, regardless of the modeled device, is sufficiently accurate when the voltage lies between 0.9 p.u. and 1.1 p.u. (the yellow zone). In our work, according to constraint (8), which is used for all optimizations, the voltage is constrained in the range [0.95, 1.05] p.u. (highlighted in green in the figure), which implies that the ZP approximation is valid for our setting.

\begin{figure}
    \centering
    \includegraphics[width=0.8\linewidth]{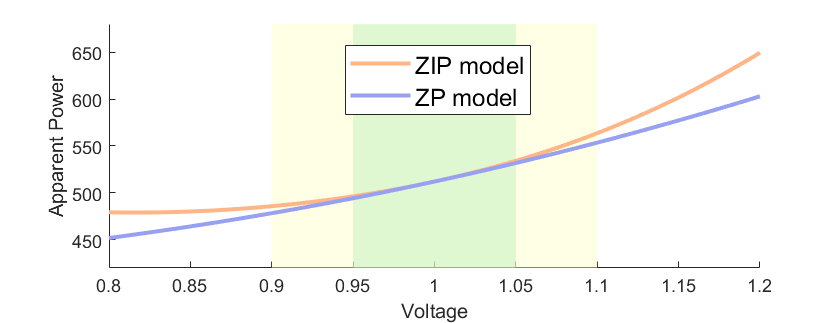}
    \caption{\Sajjad{Power demand dependence of an air conditioner on voltage in accordance with ZIP and ZP models. The green zone illustrates the voltage constraints in this research, and the yellow zone is the accurate region of the ZP approximation}}
    \label{zipvszp} 
\end{figure}}
\subsection{Power Flow Equations} 
\subsubsection{Branch Flow Model}
The branch flow model encapsulates the complete AC power flow, with the equations describing the system state as follows \cite{farivar2013branch}: 
\begin{equation} 
    \sum_{k:i\rightarrow k}s_{i,k}  = s_{\pi_i, i} - z_{\pi_i,i}\ |I_{\pi_i, i}|^2 - s_i,
    \label{BFM}
\end{equation}
where $v_{\pi_i} - v_i  = z_{\pi_i, i}\ I_{\pi_i, i},$ $s_{\pi_i, i}  = v_{\pi_i}\ I_{\pi_i, i}^*$. Note that superscript $(\cdot)^*$ denotes the conjugate of a complex number.
\subsubsection{Linearized Distribution Flow}
Linearized distribution flow (LinDistFlow) simplifies the branch flow model described in \eqref{BFM} by neglecting branch power losses \cite{baran1989optimal} and is widely adopted in several DN studies. The power flow equations under this model are given by $\sum_{k:(i,k)\in\mathcal{L}}p_{ik} = p_{ji} - p_i$, $\sum_{k:(i,k)\in\mathcal{L}}q_{ik} = q_{ji} - q_i$, as active and reactive power balances and $v^2_i = v_j^2 - 2(r_{ij}p_{ji} + x_{ij}q_{ji})$ the equation for finding subsequent voltage profile.
 \subsection{Reconfiguration of DN} \label{reconfiguration} 

\Sajjad{Reconfiguration of DNs is a well-studied topic in which the topology of the networks is changed by turning existing line switches on/off \cite{taylor2012convex}. The existence of switches in lines, which can also be used to isolate potential faults in lines, provides a paramount number of possible configurations for the grid. This typically occurs to reduce power loss, balance the network, rectify the voltage profile, enhance network restoration, and improve network reliability \cite{mahdavi2023robust, li2023restoration}. The fact that the operator can perform the changes in topology swiftly, only by sending on/off commands to switches, makes this capability a great fit for responding to a potential LAA.}

{In this subsection, a set of mixed-integer linear programming (MILP) constraints is introduced primarily to determine the configuration that maintains the nodal voltages closest to their nominal values. 
\Sajjad{We integrate the ZP approximation model into the power flow constraints (which are based on LinDistFlow model) to capture their voltage dependency.} As we show in Section \ref{results}, these approximations result in \Sajjad{formulating an MILP for the network reconfiguration problem} and provide significant computational speed-ups compared to the mixed-integer second-order cone programming (MISOCP) one.

$\bullet$ \textit{Connectivity Constraints:} {This set of constraints determines the connection between the nodes while keeping the overall graph radial. For brevity, we have omitted these constraints here, and they can be found in \cite{taylor2012convex}.}

$\bullet$ \textit{Power Flow Constraints:}
Below, we present the optimization problem's power flow constraints, which are taken from the DistFlow and ZP models:
\begin{equation}
    |p_{ij}|,|q_{ij}|  \leq Mb_{ij}, \label{p-z}
\end{equation} 
\begin{equation}
    \sum_{j:(i,j)\in \mathcal{L}} p_{ij} = p_i^f, \sum_{j:(i,j)\in \mathcal{L}} q_{ij} = q_i^f, \hspace{1cm} i \in \mathcal{N}^f \label{P_flow}
\end{equation} 
\begin{equation}
    \sum_{j \in \mathcal{N}} p_{ji} - p_{ij} = p_i^{\text{zp}}, \sum_{j \in \mathcal{N}} q_{ji} - q_{ij} = q_i^{\text{zp}}, \hspace{3mm} i \in \mathcal{N}\setminus \mathcal{N}^f \label{p_flow}
\end{equation} 
\begin{equation}
    s^{\text{zp}}_i(v_i) = p^{l}_i (\alpha'_p  + \gamma'_p v_i^2) + j q^{l}_i (\alpha'_q + \gamma'_q v^2_i),
 \label{ZP_loads}
\end{equation}
\begin{equation} 
    \underline{v_i}^2 \leq v_i^2 \leq \overline{v_i}^2,
    \label{Vlims}
\end{equation} 
\begin{equation}
    \hat{v}^2_{ij} \leq Mb_{ij}, \label{aux_V1_socp}
\end{equation} 
\begin{equation}
    \hat{v}^2_{ij} \leq v^2_i - 2 (r_{ij}p_{ij} + x_{ij}q_{ij}), \label{auxV2}
\end{equation} 
\begin{equation} 
    v^2_i = \sum_{j \in \mathcal{N}} \hat{v}^2_{ji}. \hspace{1cm} i \in \mathcal{N}\setminus\mathcal{N}^f
    \label{main_v}
\end{equation} 
Equations \eqref{p-z} - \eqref{p_flow} represent the power flow constraints, \eqref{ZP_loads} is ZIP load constraint, and \eqref{Vlims} - \eqref{main_v} are voltage constraints. The auxiliary variable of $\hat{v}$ makes the optimization follow disciplined convex programming rules in Python. Additionally, we consider all the lines to have switches.

 \Sajjad{\subsection{Inverter-Based DER}\label{DER-preliminary}
 
Inverter-based DERs can adjust the ratio of active to reactive power in their output. We use the following constraint to model such generations in DNs:

\begin{equation}
    \sqrt{p_{\text{der},i}^2 + q_{\text{der},i}^2} \leq \overline{s_{\text{der},i}} \label{der_limit}
\end{equation}

Note that, to integrate DERs into the power flow model, we consider them as negative loads.}
\subsection{Load-Altering Attack Model}\label{attack_model} 
Some major manufacturers of high-wattage IoT-controllable devices have acknowledged the presence of security vulnerabilities in their products. The concept presented in the LAAs involves attackers leveraging these vulnerabilities to transform a group of such devices into bots and toggle them on and off. \Sajjad{This coordinated action is designed to disrupt the system's stability.} Based on this, power balance equations change into 
\begin{equation} 
\begin{cases}
     \sum_{j \in \mathcal{N}} x_{ji} - x_{ij} = x_i \hspace{1.3 cm} i \in \mathcal{N}\setminus \mathcal{N}^f , i\notin \mathcal{N}^a,\\
     \sum_{j \in \mathcal{N}} x_{ji} - x_{ij} = x_i + x^a \hspace{0.4 cm} i \in \mathcal{N}\setminus \mathcal{N}^f , i\in \mathcal{N}^a,
     \label{p_LAA_distflow}
\end{cases}
\end{equation}

where $x$ is a short-hand notation, with $x = p$ for active power and $x = q$ for reactive power. \Sajjad{LAA differs from normal load deviations in cause, timing, scale, and detectability. Unlike random forecasting errors or unforeseen events, LAA is adversary-driven, strategically coordinated, and deliberately timed to exploit vulnerable grid nodes, often exceeding typical variability bounds. Their targeted and synchronous nature makes them more disruptive and distinguishable from ordinary, statistically distributed load swings.}

In the following, based on the provided models and context, we first analyze the impact of LAA on DNs. Then, we propose a game-theoretic mitigation scheme.

\Sajjad{\subsubsection{EVCS Load Altering Attack}

As detailed in \cite{maleki2025survey}, multiple devices, including EV chargers, air conditioners, heat pumps, inverters, and LED TVs, could be compromised as LAA targets. To provide an example of the attack path, we use the specific case of LAA, in which the attacker targets EVCS. 
Several common vulnerabilities and exploits (CVEs) have been identified in EV chargers \cite{soleymani2024data}. For instance, CVE-2024-43659 details a vulnerability in a commercial AC EV charger that allows an attacker to obtain default credentials from the firmware. According to this CVE, access to a charger’s firmware exposes the file containing the default credentials shared across all chargers of this model. This is a critical issue because many users do not change the default password, granting an attacker high-privilege access. This vulnerability provides the opportunity to control a large number of these devices by executing false commands. Synchronized execution of these commands on chargers in a concentrated region results in a successful LAA. }

\section{Effects of LAA on DNs}\label{effects} 
{In this section, we analyze the impact of LAA on DNs. Our objective is to derive \emph{closed-form expressions} for the voltage profile of the network with \Sajjad{voltage-dependent} loads under LAA and for the minimum compromised devices required to cause nodal voltage safety violations. Note that the grid's voltage under LAA can also be computed by solving the power flow equation \eqref{BFM} through an iterative approach such as the backwards-forward sweep (BFS) technique. However, unlike closed-form equations (which we derive in this section), the iterative techniques do not yield analytical insights into the impact of LAA. Furthermore, the closed-form expressions obtained in this section are crucial for designing defense strategies to mitigate LAA.
\subsection{Closed-form Approximation of Nodal Voltages} \label{closed_form} 
To derive the closed-form expressions for the system voltages under LAAs, we make two approximations: (i) employing LinDistFlow formulations and (ii) utilizing the ZP model.
\subsubsection{Without LAA}\label{closed-noLAA}
First, we model the DN without LAA and analyze the power flow equations.
Integrating \eqref{ZP} into voltage equation results in 
\begin{equation} 
   v_k= \sqrt{v_{1}^2 - 2\sum_{i \in \mathcal{D}_k} \left(r_{\pi_i, i}p^{\text{zp}}_{\pi_i, i} + x_{\pi_i, i}q^{\text{zp}}_{\pi_i, i}\right)}, \label{general_v}
\end{equation}
where $ p_{\pi_i, i}^{\text{zp}} =  p_{\pi_i, i}(\alpha'_{p} + \gamma'_{p} v_i^2),$ $ q_{\pi_i, i}^{\text{zp}} =  q_{\pi_i, i}(\alpha'_{q} + \gamma'_{q} v_i^2).$ Next, we perform a variable change $(u_k=v_k^2),$ which results in a set of linear equations, which can be written in matrix form as $\textbf{U}_{(N-1)\times 1} = \Omega_{(N-1)\times N}\begin{bmatrix}
        1\\
        \textbf{U}
    \end{bmatrix}_{N \times 1}$, where $\textbf{U}$ is the vector of squares of voltages, and $\Omega_{(n-1)\times n}$ is the matrix with entries: 
\begin{equation}
\omega_{i, 1} = 1- \sum_{m \in \mathcal{D}_i} (2r_{\pi_m, m} p_{\pi_m, m}^0 \alpha'_{p} +2x_{\pi_m, m} q_{\pi_m,m}^0 \alpha'_{q}),
\label{w}
\end{equation} 
\begin{equation}
\omega_{i, k} = 
\begin{cases}
\sum_{c=2}^i -2 r_{\pi_c, c} p_k^0 \gamma'_{p} -2 x_{\pi_c, c} q_k^0 \gamma'_{q}, \ \ \text{if} \hspace{0.2cm} i \in \mathcal{D}_{k}\\
\omega_{\pi_i, k},  \ \  \text{otherwise},
\end{cases}
\label{w'} 
\end{equation}
where $2\leq i,k \leq N$ ($N$ is number of buses). We rewrite the system of linear equations as 
    $(\textbf{I}_{(N-1) \times (N-1)} - \Omega'_{(N-1) \times (N-1)}) \ \textbf{U}_{(N-1)\times 1} = \Omega''_{(N-1) \times 1}$, in which  $\Omega''_{(N-1)\times 1} = [\omega_{2,k}]$, $\Omega'_{(N-1) \times (N-1)} = [\omega_{i,k}]$ for $i = \{3,4,...,N\}$, and $k \in \mathcal{N}$.

{\subsubsection{With LAA}\label{closed-laa} \label{closed_form_LAA}
Here, we analyze the voltage profile of the network under LAA. For this, we integrate the introduced LAA in Section \ref{attack_model} into \eqref{general_v} and obtain:
\begin{equation}
    v_{k}^a =  
    \sqrt{v_{1}^2 - 2\Delta_k - 2p^{A}_ar_{k,a} - 2q^{A}_ax_{k,a}}, \label{general_v_LAA} 
\end{equation}
in which $\Delta_k = \sum_{i \in \mathcal{D}_k} \left(r_{\pi_i, i}p^{\text{zp}}_{\pi_i^, i} + x_{\pi_i, i}q^{\text{zp}}_{\pi_i, i}\right)$, \ $r_{k,a} = \sum_{i \in \{\mathcal{D}_a \cap\mathcal{D}_k\}} r_{\pi_i, i},$ and $ x_{k,a} = \sum_{i \in \{\mathcal{D}_a \cap\mathcal{D}_k\}} x_{\pi_i, i}$. \Sajjad{Further details of obtaining \label{general_v_LAA} and the rationale behind it are provided in Appendix \ref{appendix}.} This change results in a new set of coefficient matrices. To calculate the attacked system's square of voltages vector $(\textbf{U}^A)$, we solve
$\textbf{U}^{A}_{(N-1)\times 1} = \Omega^{A}_{(N-1)\times N}\begin{bmatrix}
        1\\
        \textbf{U}^{A}
\end{bmatrix}_{N \times 1}$. To obtain the coefficient matrices, \eqref{p_LAA_distflow} is dragged into LinDistFlow as the ZP model is imposed on them. The final results are $\omega_{i,k}^A = \omega_{i,k} + \omega_{i,k}^a$, for $i\geq 2$ and $k\geq 1$; $\Omega^{A}_{(N-1)\times N} =[\Omega^{A''}_{(N-1)\times 1} \ \Omega^{A'}_{(N-1)\times (N-1)} ]$, in which for $i\geq 2$ and $k\geq 2$ and $\omega_{i,1}^a = \sum_{c \in\{\mathcal{D}_i\cap\mathcal{D}_a\}} -2p_a^{A_{0}} \alpha'_{p} r_{\pi_c, c} - 2q_a^{A_{0}} \alpha'_{q} x_{\pi_c, c}$ , $\omega_{i,k}^a =
\begin{cases}
   -2r_{\pi_i,i}p^a\gamma'_{p} -2x_{\pi_i,i}q^a\gamma'_{q},  \ \ \text{if} \hspace{0.2cm}i\in \mathcal{D}_{a}\\
    \omega_{\pi_i,k}^A, \ \  \text{otherwise}.
\end{cases}$

If there is more than one attacked node at a time, the impacts will be summed up. As a result, the new resulting coefficients are: $\omega^{a_t}_{i,j}=\sum_{a \in \mathcal{A}} \omega^a_{i,j}, \ \forall i\in [2,N] \ \& \ j \in \mathcal{N}$.
{Subsequently, $\omega_{i,k}^A = \omega_{i,k} + \omega_{i,k}^{a_t}$, for $i\geq 2$ and $k\geq 1$.}

\Sajjad{Note that when there is an attack in a leaf bus (the last bus of each branch), $r_{k,a}$ and $x_{k,a}$ have the highest possible values. As a result, the voltage drop resulted from $(- 2p^ar_{k,a} - 2q^ax_{k,a})$ in \eqref{general_v_LAA} is higher and obtained voltages shrink. In conclusion, attacks on the leaf buses yield the most detrimental effects.} \Sajjad{However, this conclusion only holds when there are no DERs in the system. Since DERs can be modeled as negative loads, the direction of power flow on some lines may change, and the voltage drop trend may not be the same.}
\subsection{Analytical Insights into the Attack Impact}\label{critical_attack_cform}
\Sajjad{Using the results obtained in Subsection \ref{closed_form}}, we further obtain the minimum number of attacked devices, which leads to voltage safety violations.
We call such a threat the ``\textit{critical attack}". For this, we consider the voltage of the leaf bus as a known variable ($v_{th}$). The new unknown variable is $p^a$, and based on the attacked device type, we can find $q^a$ via $q^a=\frac{q_d}{p_d}\ p^a$. So the new set of coefficients for obtaining voltages of buses except for the leaf one and the active power of the critical attack is forming $\Omega^d$ ($\Omega^{d} = [\Omega^{d''} \Omega^{d'}]$) in which 
\begin{align}
    \omega_{i,1}^{d} & = 
    \begin{cases}
       \omega_{i,1} + v_{th}^2\omega_{i, a}, \ \ \text{if}\ i \neq a,\\
        \omega_{i,1} + v_{th}^2(\omega_{i, a}- 1), \ \ \text{if}\ i = a,    
    \end{cases} \\[2pt]
\omega_{i,k}^{d} & = 
\begin{cases}
    \sum_{c\in \mathcal{D}_i} -2r_{\pi_c,c}\alpha'_{p}-2 \frac{q_{d}}{p_{
    d}}x_{\pi_c,c}\alpha'_{q}, \ \ \\
    \quad \quad \quad \ \text{if}\  k=a, i\in\mathcal{D}_a,\\
    \omega_{\pi_i,k}^{d}, \ \ \text{if} \ k=a, i\notin\mathcal{D}_a,\\
    \omega_{i,k}, \ \ \text{otherwise}.
    \end{cases} 
\end{align}
Hence, we can solve the linear system of equations given by $(\textbf{I}^{a} - \Omega^{d'})\textbf{X} = \Omega^{d''}$, in which $\textbf{X}$ is a vector with the same dimension as $\textbf{U}$ where all elements of it are the same as $\textbf{U}$ except for one in which $\textbf{X}$ contains $p^a$ instead of $u_a$ (since we already know $u_a = v_{th}^2$). Additionally, $\textbf{I}^{a}$ is the identity matrix except for the element $(a, a)$ which equals $0$. This gives the bus voltages as well as the required $p^a$. Then, we can find the number of required devices in bus $n$ using $c(n) = \frac{p^a}{p_d}$.

The closed-form expressions provided in this section are not only used to identify the worst effect of the LAAs, but also inspire our mitigation method to find the optimal defensive action in the following sections. Additionally, the introduced ``critical attack" has been implemented in finding the optimal action of the strategic attacker later in this paper.
\section{Mitigating LAA via Reconfiguration}\label{LAAmitigation} 
In this section, we introduce a novel technique to mitigate LAAs by reconfiguring the DN topology \Sajjad{and adjusting the outputs of DERs}. We exploit a sequential game-theoretic interaction in which, following the LAA launched by the attacker, the defender reconfigures the network to react optimally to the threat. \Sajjad{But first, we clarify the position of this work in the defense process and cyber-resilience of power grids.

\subsection{Defense Steps in Power Grid and Position of This Paper}
Cyber resilience in power system involves three stages \cite{liu2025cyber} -- 
(i) \emph{pre-attack stage} that involves identifying potential attacks and taking preventive measures, (ii) \emph{during-attack stage}, which includes attack detection and mitigation, and (iii) \emph{post-attack stage}, which involves locating and isolating the compromised system. 
In an LAA scenario, a group of devices are controlled by the adversary (e.g., botmaster). As a result, even if the operator is able to locate the attack, it may not be possible to isolate the loads quickly. 
The proposed method thus falls under the category of \emph{attack mitigation} (i.e., \emph{during-attack stage}), which can be viewed as a quick response to the attack to keep the grid's voltage in the desired range. This provides more time for the \emph{cyber-recovery} step, which includes exactly locating, isolating, and physically removing the attack (which typically incurs additional delays).
\begin{figure}
    \centering
    \includegraphics[width=0.85\linewidth]{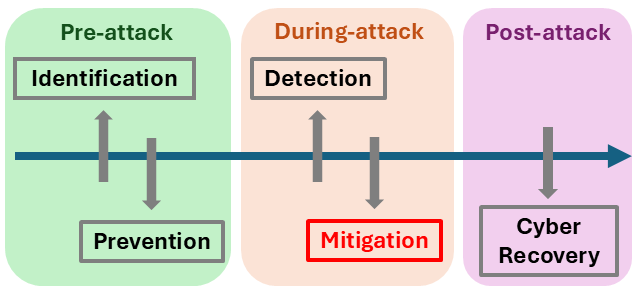}
    \caption{\Sajjad{Defense timeline against LAA.}}
    \label{defense-timeline}
\end{figure}
}
\subsection{Mitigation Design and Intuition} 
The intuition behind the proposed defense technique of reconfiguring the DN lies in the analytical insights derived in Section \ref{effects}. Based on our analysis, LAAs targeting the leaf buses of the DN lead to the greatest attack impact (i.e. deviation of the voltage from the nominal values).
In this context, reconfiguring the DN changes the position of the leaf buses, thus alleviating the attack impact. \Sajjad{The presence of DERs in the grid challenges the notion that LAAs on leaf nodes always have the most detrimental impact. However, we utilize inverter-based DERs as an additional tool to mitigate the effects of LAAs. In our approach, when inverter-based DERs are present, they adjust their active and reactive power outputs to help maintain an optimal voltage profile across the grid under LAA conditions.}

It is worth noting that the proposed mitigation leverages the \emph{pre-existing} capabilities of the DN (e.g., devices enabling network reconfiguration are primarily installed to reduce power losses and/or voltage deviations) and, hence, does not require new infrastructure.
Furthermore, in the proposed scheme, the system will be reconfigured only when an attack is detected, \Sajjad{thus avoiding unnecessary defensive actions} (note that cyberattacks are somewhat rare events). 
The attack detection module can be based on existing model-driven or data-driven approaches for detecting LAAs. The reader can refer to past works, including \cite{jahangir2023deep} and \cite{li2022adaptive} in this area for more details.
\subsection{Stackelberg Game for Attack Mitigation}

\Sajjad{We adopt the common cybersecurity practice of conservatively assuming a strong adversary with full system knowledge. This assumption enables us to evaluate the worst-case impact of cyber-physical attacks on DN operations and DERs. Although this model represents an upper bound in terms of adversarial capability, it is consistent with established practice in the security literature, where strong attackers are assumed to stress-test system vulnerabilities and provide benchmarks for resilience \cite{ghosh2024bi, lakshminarayana2021moving, sanjab2016data}.

In practice, the gap between realistic attackers and an omniscient adversary is often smaller than presumed. Due to the high degree of digitization and data availability in modern power systems, an attacker can gain significant visibility into system states. Grid topology, historical demand and generation profiles, market data, and renewable generation forecasts are frequently accessible through public datasets, regulatory filings, or system probing \cite{acharya2020public}. Moreover, insider threats, compromised DER controllers, and advanced reconnaissance techniques can provide attackers with near-complete knowledge of system conditions.

By adopting a strong attacker model, we ensure that the results capture the maximum achievable disruption and highlight systemic vulnerabilities that may otherwise be overlooked under weaker assumptions. Importantly, if the system is resilient to an omniscient adversary, it will inherently be robust against less capable attackers. Conversely, defensive strategies developed under weaker threat models may fail to provide sufficient protection when adversaries obtain more information or resources than anticipated. Thus, the strong-adversary model is not only a theoretical abstraction but also a practical safeguard, ensuring that mitigation strategies address the full spectrum of plausible threats.

We model the strategic interaction between the attacker and the defender as a non-cooperative Stackelberg game. This framework is the natural mathematical framework modeling an asynchronous, sequential decision process between two opposing and strategic agents \cite{shukla2022robust}, where the attacker acts first by choosing an LAA strategy. Subsequently, the defender optimally reacts by reconfiguring the system to mitigate the attack. The non-cooperative framework allows us to consider explicitly the worst-case scenario, which is essential in deriving robust defenses. Table \ref{game_assumptions} summarizes the game's assumptions and details.

In our problem, we assume that complete information about the game is available to both players. Therefore, Bayesian games are not a suitable fit, as they are intended for modeling incomplete information in non-cooperative games \cite{wu2021agent} and introduce significant complexity. The only uncertainty considered in this work, related to the attack location, is incorporated through a probability distribution directly in the function of the defender, as detailed later in this section. Moreover, since cyberattacks on power grids are relatively infrequent events, we do not consider iterative methods such as reinforcement learning ones, which would require frequent interactions between the attacker and the defender.}
 \begin{table}[t]
     \caption{\small Non-cooperative game assumptions.} 
     \centering
     \begin{tabular}{|c|>{\centering\arraybackslash}p{3.3cm}|>{\centering\arraybackslash}p{3cm}|} \hline
           Player& Attacker & Defender  \\ \hline
          Role & Leader & Follower \\ \hline
         \multirow{2}{*}{Knowledge} & Potential defensive reactions, Grid topology  & Attack strategy, Grid topology \\ \hline
          \multirow{3}{*}{Actions} & \multirow{3}{*}{Switching loads} &  Changing the topology \& Adjusting DERs output (if any exist)\\ \hline
          \multirow{2}{*}{Goal} & Max. voltage constraint violation & \multirow{2}{*}{Min. attack impact} \\ \hline
     \end{tabular}
     \label{game_assumptions}
         \vspace{-1em}
 \end{table}

A Stackelberg game with two players consists of a leader and a follower. The leader always commits first to maximize their own objective function by anticipating the follower's reaction. Then, given the action of the leader, the follower picks their best (i.e., optimal) response to maximize their own objective function. Since LAAs are rare incidents in the network, we propose a reactive mitigation method. The proposed Stackelberg game can be defined as $\mathcal{H} = \{(A,D), (\mathcal{S}_A, \mathcal{S}_D), (F_A, F_D)\}$, in which $A$ and $D$ are the players (attacker and defender), $\mathcal{S}_A, \mathcal{S}_D$ denote sets of actions of players, and $F_A, F_D$ denote their rewards.

The attacker's set of actions, $\mathcal{S}_A$, is launching LAA in any of the buses (one or two buses at a time by assumption). The set of defense actions is the set of all possible reconfigurations discussed in Section \ref{reconfiguration} such that $\mathcal{S}_D = \{1,2, \hdots, N_B\}$ denotes the set of indices of all possible reconfiguration matrices $B = [b_{ij}]_{1\leq i,j\leq N} \in \{0,1\}^{N \times N}$ whose entries meet the connectivity constraints and $N_B$ represents the number of such matrices. Henceforth, we denote by $B(d) = [b_{ij}(d)]_{1\leq i,j\leq N}$ the adjacency matrix of the network as a result of a specific defense $d\in \mathcal{S}_D$.
The attacker aims to maximize the voltage deviation, as a result, we define $F_A(d,a) = \sum_{n \in \mathcal{N}} \left |v^2_{nom} - v^2_{n}(d,a)\right |$, as the attacker's objective function. 


 
 The defender's objective is to reconfigure the system to minimize the square of the voltage deviation above. Achieving this goal with minimal changes can decrease the maintenance requirements for switches and reduce the likelihood of switching failures. To take this factor into account, we add a penalty term and the resulting reward function of the defender is 
\begin{equation} 
    F_D^{perf} = -\sum_{n \in \mathcal{N}} \left |v^2_{nom} - v^2_{n}(d,a)\right | - pen(d),
    \label{obj_pen} 
\end{equation}
where $pen(d) =  \sum_{i=1}^{N}\sum_{j=1}^{N}\left|b^{pre}_{ij}-b_{ij}(d)\right|$ is the penalty term for enforcing a system reconfiguration with the minimum switches possible. $F_D^{perf}$ is relevant when the defender is capable of perfect attack localization. However, due to noisy measurements, the defender might not be able to do this. Instead, we assume that the defender is only able to locate a \Sajjad{neighborhood} of the attack, i.e., a connected cluster of buses which contains the bus under attack. We further assume that the defender has a favourite candidate bus under attack, denoted by $n_{att}$, but does not discard the attack possibilities of other buses in the located cluster. To model this, we define a discrete probability vector $\sigma = [\sigma_1, \sigma_2, \hdots, \sigma_N]$ with entries: 
 \begin{equation} 
     \sigma_\ell = 
     \begin{cases}
         \rho, & \textbf{if} \hspace{0.3cm} \ell = n_{att},\\
         \frac{1-\rho}{|\mathcal{A}_{n_{att}}|}, & \textbf{if} \hspace{0.3cm} \ell \in \ \mathcal{A}_{n_{att}},\\
         0, & \textbf{otherwise}.
     \end{cases}
     \label{probability}
 \end{equation}
Above, $\sigma_\ell$ represents the likelihood that the defender assigns to bus $\ell$ being under attack such that $\sigma_\ell \in[0,1]$ and $\sum_{\ell \in \mathcal{N}} \sigma_\ell = 1$. Additionally, $\rho \in [0.5, 1]$ denotes the likelihood of the defender's favourite candidate. The subset $\mathcal{A}_{n_{att}}$ is the set of buses directly adjacent to $n_{att}$, along with the buses directly adjacent to those; all these buses are considered as the other potential candidates by the defender. Their likelihood is the remaining probability $1-\rho$ split equally between the $|\mathcal{A}_{n_{att}}|$ other candidate buses. 


Taking this uncertainty (of \Sajjad{precisely detecting} the attack location) into account at the defender's results in the following reward: 
\begin{equation} 
    F_D (d,a)= -\sum_{\ell\in\mathcal{N}}\sigma_{\ell}\sum_{n\in\mathcal{N}} \left |v^2_{nom} - v^2_{n}(d,a_{\ell})\right | - pen(d),
    \label{Obj_def}
\end{equation}
which represents the expected reward over this uncertainty. Obtaining the optimal attack and defense requires computing the Stackelberg equilibrium. As discussed earlier, in this paper, the attacker commits the attack first and then the defender reacts. 
\begin{definition}
    The best response of the defender to an action $a \in \mathcal{S}_A$ is defined as:
    \begin{equation}
        r(a) = \arg\max_{d\in\mathcal{S}_D} F_D(d,a). \label{r_a}
    \end{equation} 
\end{definition} 
\begin{definition}
    A profile of actions $(a^*, d^*) \in (\mathcal{S}_A, \mathcal{S}_D)$ is a Stackelberg equilibrium iff 
\begin{equation} 
\begin{cases}
    F_A(r(a^*), a^*) & \geq  F_A(r(a), a), \hspace{0.7cm}\forall a \in\mathcal{S}_A\\
    \hspace{1.8cm}d^* &= r(a^*).
\end{cases}
\end{equation}
\end{definition}

 Intuitively, the attacking action at the Stackelberg equilibrium is the one maximizing the attacker's reward under the defender's best reaction. Furthermore, the defender's best reaction to $a^*$ is its Stackelberg equilibrium action. A Stackelberg equilibrium is ensured to exist if the defender's optimal response exists for every attack. Assuming that normally open points exist in the distribution system, this ensures that at least one system reconfiguration is possible and that the discrete feasible set in \eqref{r_a} is non-void, leading to the existence of the solution (if the corresponding constraints are met). 
{\subsection{Bayesian Optimization}}
{BO provides algorithms for optimizing black-box functions, whose mathematical expression is unknown or too complex to analyze \cite{garnett2023bayesian}. Instead, BO relies on the function's observed values for given inputs. This also makes BO suitable for optimizing computationally expensive functions that need to be evaluated exhaustively \cite{brochu2010tutorial}. \Sajjad{BO does not have a mathematical guarantee of finding the global optimum, but can significantly reduce the complexity of analyses.}

In our case, obtaining the best response function for all possible attacks requires solving a separate optimization problem for each attack, \Sajjad{resulting} significant computational complexity. To address this issue, we consider the attacker's reward to be a black-box function and utilize BO to explore and approximate the optimal attack efficiently. {Similarly to \cite{yan2025optimised}, we first build a probabilistic model ($F^p_A$) for $F_A(a, r(a))$ using a Gaussian process $(\mathcal{GP})$. This model is based on sample attacks $\mathcal{T} = \{a^{s}_i\}$ and their corresponding $\{F_A(a^{s}_i, r(a^{s}_i))\}$, which is obtained using:
\begin{equation} 
    F^p_A \sim \mathcal{GP}(m(a), k(a,a')), \label{F_p}
\end{equation}}
{where $(m(a))$ is the mean function and $(k(a,a'))$ is the covariance kernel. Note that we sample the attack locations to form the $\mathcal{T}$ from all branches to have a good initial estimate model. Afterwards, we use the expected improvement (EI) function to pick the next attack action for evaluation via:
\begin{equation} 
    a^{\mathrm{next}} = \arg\max_{a \in \mathcal{S}_A} EI(a), \label{a_next}
\end{equation}
and add this new point to the $\mathcal{T}$ and update $F^p_A$. The EI function is $EI = \mathbb{E}[\max\{0, F_A^{\max} - F^p_A(a)\}]$. The process is repeated until the stopping criteria are met.}


Algorithm \ref{alg_stack_game} describes the method of finding the players' actions at the Stackelberg equilibrium.
\begin{algorithm}


    \KwData{$\mathcal{H} \ , \ \sigma = \{ \sigma_1, ... , \sigma_N\}$}
    \KwResult{$r(a^*)$ , $a^*$}
    \begin{algorithmic}[1]
    \STATE Sample $\{a^i\}_{i = 1}^{n_0}$ from $\mathcal{S}_A$, form the set $\mathcal{T}$, and compute $r(a), \ \forall a\in \mathcal{T}$
   {by solving}
   \hspace{1.5em} $r(a) = \arg\max_{d\in \mathcal{S}_D} F_D(d,a)$ s.t. connectivity and power flow constraints as an MILP optimization problem via SCIPY and CVXPY;
    \STATE Compute $F_A(r(a),a), \ \forall a \in \mathcal{T}$;\\
    \STATE Construct a probabilistic model of the $F_A(r(a),a)$ as in \eqref{F_p};\\
    \STATE Find the next query point $(a^{\mathrm{next}})$ using \eqref{a_next}, attach it to $\mathcal{T}$, and compute $F_A(r(a^{\mathrm{next}}),a^{\mathrm{next}})$;\\
    \STATE Repeat until the stop condition is met;\\
    \STATE Choose $a^* = \arg\max_{a\in \mathcal{T}}F_A(r(a),a)$;\\
    \STATE Compute $r(a^*) = \arg\max_{d\in \mathcal{S}_D}F_D(d,a^*)$ as in step 1; \\  
    \caption{Computing the Stackelberg equilibrium using BO.}
    \label{alg_stack_game}
    \end{algorithmic}
\end{algorithm}

 The constraints of the reconfiguration optimization (as in steps one and seven of the algorithm \ref{alg_stack_game}) are linear in the square of the voltages, hence, justifying our choice of the distance between the squares of voltages in the reward function, leading to a linear program instead of a quadratic one (obtained by a variable change $u_i = v_i^2$). The resulting MILP optimizations are carried out in Python; the optimization modeling language is CVXPY, and the solver is SCIPY. We find the reward of the attacker for $(r(a), a),\ \forall a \in \mathcal{T}$. Finally, the action corresponding to the maximum of $F_A(r(a), a)$ is selected as the attacking strategy in our Stackelberg formulation. Additionally, $r(a^*)$ corresponding to the Stackelberg attack action $a^*$ is the optimum defensive strategy at the Stackelberg equilibrium: $d^* = r(a^*)$.

\subsection{Resource-Constrained Attacker} \label{min_effort_att} 
Drawing from our discussions in Section \ref{critical_attack_cform}, each bus has a distinct ``critical attack" leading to voltage constraint violation. Given the attacker's tendency for launching such a ``critical attack", their potential action will not only occur across different buses but also vary in magnitude.

To accommodate this feature, we define a new game $\mathcal{H}' = \{(A,D), (\mathcal{S}_A, \mathcal{S}_D), (F'_A, F_D)\}$ in which the attacker launches the ``critical attack" which we call them the ``resource-constrained attacker". Note that the critical attack in each bus is manipulating the minimum devices ($c(n)$) in that bus to cause a voltage constraint violation. Indeed, $c(n)$ for each attack is unique and varies with the attack location.
In this modified game, $F'_A$ comprises two components: the total nodal voltage deviation and the attack magnitude. The attacker seeks to maximize the former while minimizing the latter. However, these two terms cannot be simply summed due to their disparate physical characteristics. Therefore, we propose the following reward: 
\begin{equation} 
  F'_A(d,a) = (1 - \lambda)F_A^{\mathrm{norm}}(d,a) -  \lambda c^{\mathrm{norm}}(a),  
\end{equation}
in which $0\leq\lambda\leq1$, $F_A^{\mathrm{norm}}(d,a) = \frac{F_A(d,a)}{\sum_{i \in \mathcal{S}_A}F_A(d,i)}$, and $c^{\mathrm{norm}}(a) = \frac{c(a)}{\sum_{i \in \mathcal{N}^L}c(i)}$. The parameter $\lambda$ trades off between the two components of the objective function. If $\lambda = 0$, the attacker only cares about maximizing the harm caused in the voltage profile; and if $\lambda = 1$ the attacker only cares about minimizing the attacked devices. The rest of the components of $\mathcal{H}'$ are the same as $\mathcal{H}$. The process of computing the Stackelberg equilibrium is similar to $\mathcal{H}$, and we only need to plug in the attacker's new objective function $F'(d,a)$.

\Sajjad{\subsection{Integration of inverter-based DERs}
Integration of DERs in DNs is inevitable. As a result, we also include them in our models and deploy them as another resource for alleviating the LAA impact. These resources have a capped generation capacity; however, by adjusting their power factor, the shares of their active and reactive output power can be varied. We utilize this flexibility, in coordination with network reconfiguration, by adding the constraint from \eqref{der_limit} (in Subsection \ref{DER-preliminary}) and by replacing \eqref{ZP_loads} with \eqref{der_int}.

\begin{equation}
    \begin{cases}
        \begin{split}
            s^{\text{zp}}_i(v_i) = \big[p^{l}_i (\alpha'_p + \gamma'_p v_i^2) - p_{\text{der},i}\big] & \\
            + j \big[q^{l}_i (\alpha'_q + \gamma'_q v^2_i) - q_{\text{der},i}\big],
        \end{split}
        & \ \text{if} \  i \in \mathcal{N}^{\text{DER}} \\
        s^{\text{zp}}_i(v_i) = p^{l}_i (\alpha'_p + \gamma'_p v_i^2) + j q^{l}_i (\alpha'_q + \gamma'_q v^2_i), &  \ \text{else}
    \end{cases}
    \label{der_int}
\end{equation}

}
\section{Results and Discussions}\label{results} 
\Sajjad{Here, our simulations are conducted using the IEEE 33-bus and 69-bus systems. Fig. \ref{base-configs} presents the base topologies of them, in which dashed lines are normally open (disconnected) lines (i.e., auxiliary links). There are 32 and 48 load buses in the test grids, respectively, which could be attacked. Note that the results are partially extended to the 141 bus grid in subsection \ref{141result}.} The ZIP load coefficients are set to the residential load-type F introduced in \cite{bokhari2013experimental}.
\begin{figure}
    \centering
    \color{RoyalBlue}
   \begin{subfigure}[b]{0.75\linewidth}
        \centering
        \includegraphics[width=\linewidth]{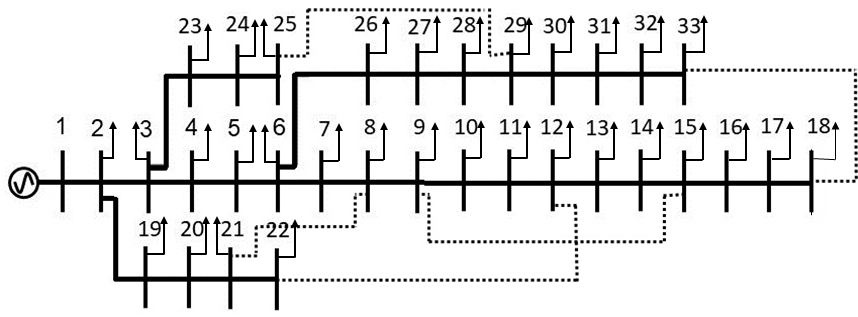}
        \caption{IEEE 33-bus grid}
        \label{33-bus}
    \end{subfigure}
    \vfill
    \begin{subfigure}[b]{0.75\linewidth}
        \centering
        \includegraphics[width=\linewidth]{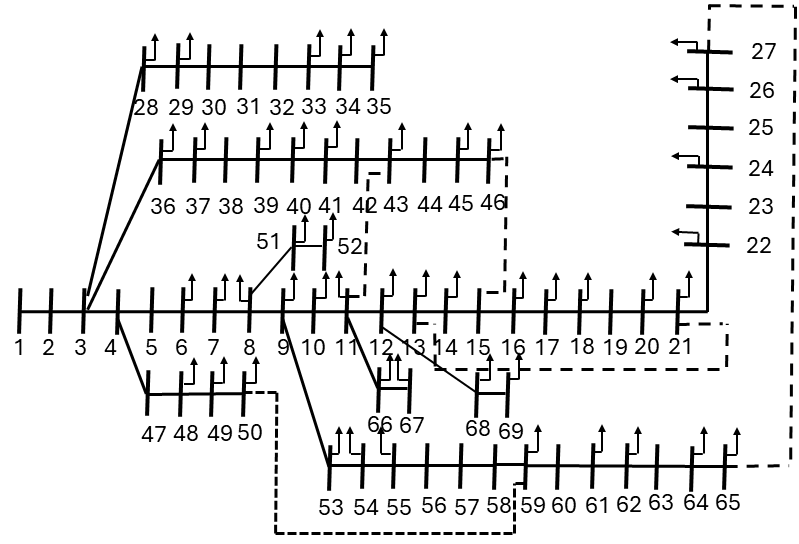}
        \caption{IEEE 69-bus grid}
        \label{69-bus}
    \end{subfigure}
    \caption{\Sajjad{Base configuration of the test cases}}
    \label{base-configs}

\end{figure}
\subsection{Critical Attack}\label{C_attack_result} 
Here, we conduct spatial analyses to determine the most effective location for launching LAAs. To quantify this, we use the load profile obtained from \cite{nyiso} for the date 05/05/2024 as the base load (without LAAs), which contains the hourly load demand in New York, US. To mimic this load profile in the 33-bus grid, we project the ratio of load changes at different hours onto the nominal load of the test network. Table \ref{c_attck_ls} presents the number of devices required for the critical attack in three of the leaf buses of the test case during the different hours of the day. These numbers are computed via the equations in Section \ref{critical_attack_cform}. We can see that the attack on the deepest bus requires fewer devices to be manipulated. This conclusion confirms our insights from Section \ref{closed_form}.
Furthermore, it also shows the dependency on the type of load (air conditioner, resistive load, etc.) and the associated ZIP load coefficients. 


\begin{table}[t]
    \caption{\small Numbers of compromised devices required to cause voltage safety violations in the 33-bus grid during different hours.}
    \begin{center}
        
    \begin{tabular}{|>{\centering\arraybackslash}p{1.3cm}|>{\centering\arraybackslash}p{1.5cm}|>{\centering\arraybackslash}p{1.4cm}|c|>{\centering\arraybackslash}p{1.5cm}|} \hline
        \multirow{2}{1.3cm}{Device\centering}  & Att. location (bus) &03:00 (Least load) & \multirow{2}{*}{09:00} & 18:00 (Peak load)\\ \hline
        \multirow{3}{1.3cm}{Air Conditioner\centering}&18& 603 & 459 & 38 \\ \cline{2-5}
        
         & 25& 5538 & 5168 & 612 \\ \cline{2-5}
         
         & 33& 1498  & 1079 & 248 \\
        \hline
         \multirow{3}{1.3cm}{Resistive heater\centering}& 18 & 256 & 134 & 17\\ \cline{2-5}
          
         & 25 & 2573 & 1582 & 186 \\ \cline{2-5}
        
         & 32 & 612 & 441 & 85 \\ 
        \hline
    \end{tabular}
    \end{center}
    \label{c_attck_ls}
\end{table}

Since the results in Table \ref{c_attck_ls} are obtained by the approximation discussed in Section \ref{effects}, we evaluate how effective these attacks are when considering the full AC power flow model. To evaluate the extent of the errors (between the voltages computed using the analytical results and those obtained from the full AC power flow model), we compare the obtained voltage profile with the results of BFS. Fig. \ref{acc_approx} shows the voltage profile calculated by the two methods during the peak load demand and the corresponding critical attack. We note that the actual errors in computing the nodal voltages using our approximations never exceed $1\%$, thereby validating the analytical results. 
 \begin{figure}[t]
      \begin{center}  \includegraphics[width=0.35\textwidth]{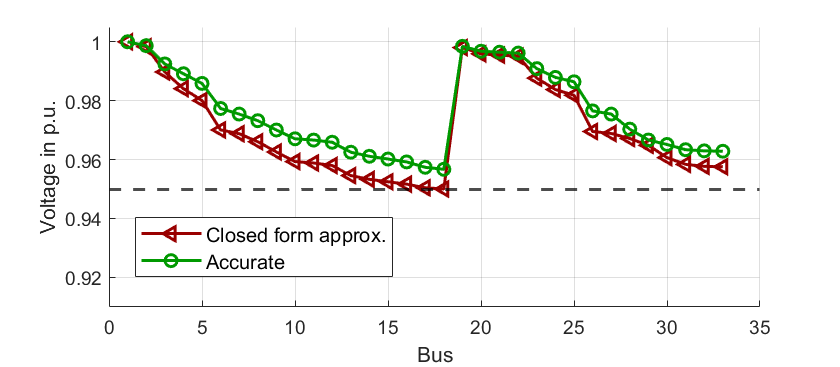} 
      \end{center}
     \caption{\small Voltage profile of the attacked (on Bus 18) 33-bus test case with the proposed closed-form equations and the accurate model.}
     \label{acc_approx}
 \end{figure} 
\subsection{LAA Mitigation} \label{LAA_mitigation_results} 
Next, we examine the proposed LAA mitigation method. {The base load profiles (without LAAs) of 33-bus and 69-bus test cases are $60\%$ and $30\%$ of their nominal load profiles in MATPOWER.} 
For scenarios i) and ii), we consider LAA attacks of magnitude $p_i^a = q_i^a = 0.30\ p.u.$ in which $i$ is the index of the attacked bus. 
\Sajjad{The} magnitude of the attack is significant enough to cause voltage safety violations in all the load buses (except those adjacent to the root) and, hence, needs to be mitigated.
Four scenarios are considered next: i) accurate attack localization, ii) errors in attack localization, iii) errors in attack localization and resource-constrained attacker, {and iv) attacks on two nodes.}

\textbf{i) Accurate Attack Localization:}
Here, we set $\sigma_{n_{att}} = 1$ and $\sigma_{\ell} = 0 \hspace{0.15cm} \forall \hspace{0.15cm} \ell \in \mathcal{N}^L\setminus \mathcal{N}^a$. Table \ref{SE_results} presents the players' actions at the Stackelberg equilibrium. Based on these results, in the 33-bus grid, the attack is launched on Bus 33. Fig. \ref{voltages} shows the voltage profiles of the grid under attack before and after reconfiguration. The results show that the proposed mitigation method is able to return the voltage profile within the constraints, hence mitigating the effects of LAA. Although the LAA on Bus 18 causes the greatest impact in terms of voltage deviations, a strategic attacker who can anticipate the defender's reaction chooses to launch the attack on Bus~33 instead to maximize their payoff. Furthermore, in the 69-bus grid, the attack is launched on Bus 27 and lines (50-59) and (58-59) change their state to reconfigure the network.

\begin{figure}[t] 
    \centering
    \includegraphics[width=0.35\textwidth]{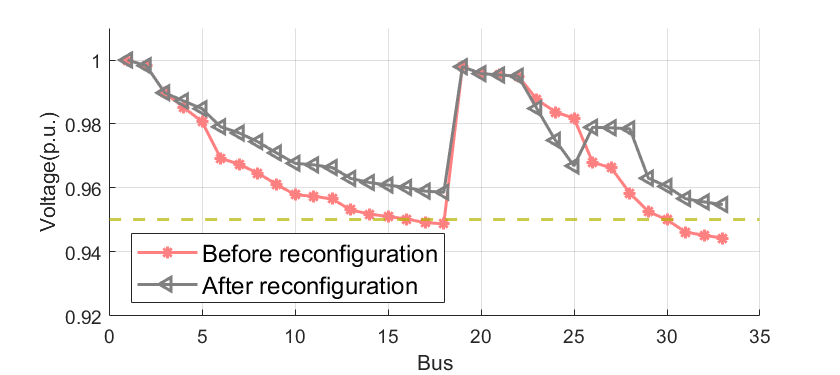} 
    \caption{\small Voltage profile of the attacked 33-bus grid before and after mitigation (reconfiguration).}
    \label{voltages}
\end{figure}

\textbf{ii) Errors in Attack Localization:}
In this scenario, following the worst-case accuracy of the detection algorithm in \cite{li2022adaptive}, we consider  $\rho = 0.7$ ($70\%$ of certainty about the location of the attack), and then the remaining $30\%$ is split equally between other buses in the neighbourhood (see Section \ref{LAAmitigation}). The results for this scenario are presented in Table \ref{SE_results}. In this scenario, the defense action should keep the voltage within the desired constraints, assuming an LAA in any of the candidate buses. This uncertainty causes a system reconfiguration that necessitates more switching. Fig. \ref{Scen_b} represents the voltage profile of the 33-bus grid after the reconfiguration if any of the three suspect buses are attacked. We can notice that the attack will be successfully mitigated and the voltage constraint will not be violated in any case. \\
Note that the players' actions for the 69-bus system are the same as in scenario i).

  \begin{figure}[t]
      \centering
      \includegraphics[width=0.35\textwidth]{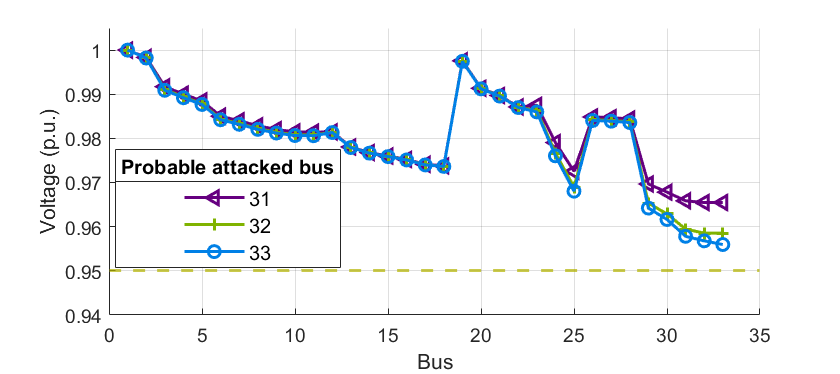}
      \caption{\small Voltage profiles of 33-bus grid attacked at any of the suspect buses after mitigation. }
      \label{Scen_b}
  \end{figure}

\textbf{iii) Resource-Constrained Attacker:}
Here, we present the resource-constrained attacker introduced in Section \ref{min_effort_att} with $\lambda = 0.5$ in their objective function.
Similar to the two previous cases, the results of this scenario can also be found in Table \ref{SE_results}. We observe that in this scenario, the attacker chooses to attack Bus 18, as the impact of load alteration on this bus will be the greatest (as the number of compromised devices is taken explicitly into account). The 69-bus case maintains the same Stackelberg equilibrium actions as the past two scenarios.

{\textbf{iv) Attacks on two nodes:}
In this scenario, with insufficient vulnerable loads, the attacker targets two locations. In the 33-bus grid, this results in affected buses 32 and 33, with a defensive response similar to Scenario ii). In the 69-bus grid, buses 26 and 27 are targeted, showing the same defensive pattern as before.} Table \ref{V_dev} represents the total voltage deviations ($\sum_{i\in \mathcal{N}}\left|v_{nom} - v_i\right|$) in the grid for different scenarios. Note that, in scenario ii), the voltage deviation of the 33-bus grid drops, but the defender needs to commit more switching, which is not desirable. Furthermore, a portion of the reduced voltage deviation in scenario iii) should be attributed to the smaller attack launched by the resource-constrained attacker.
\begin{table}[t] 
    \caption{\small Attacked bus(es) and closed/open lines at the Stackelberg equilibrium of each scenario.}
    \begin{center}
    \begin{tabular}{|c|c|c|p{0.95cm}|c|p{0.95cm}|c|} 
\hline
\multirow{2}{0.85cm}{Scenario\centering} & \multicolumn{2}{c|}{Attacked bus \centering} & \multicolumn{2}{c|}{Closed line(s) \centering} & \multicolumn{2}{c|}{Opened line(s)\centering} \\ \cline{2-7}
 &33-bus\centering & 69-bus\centering &33-bus\centering &69-bus &33-bus\centering &69-bus\\
 \hline
i\centering & \centering 33 & 27 & (25-29)  &(50-59) & (28-29) &(58-59)\\ \hline
\multirow{2}{*}{ii\centering} & \multirow{2}{*}{\centering 33} & \multirow{2}{*}{27} &(22-12), (25-29) &\multirow{2}{*}{(50-59)} &(11-12), (28-29) &\multirow{2}{*}{(58-59)}\\\hline
iii\centering & \centering 18 & 27 & (22-12) &(50-59) & (11-12) &(58-59) \\
\hline
\multirow{2}{*}{iv\centering} & \multirow{2}{*}{\centering 32\&33} & \multirow{2}{*}{26\&27} &(22-12), (25-29) &\multirow{2}{*}{(50-59)} &(11-12), (28-29) &\multirow{2}{*}{(58-59)}\\\hline
\end{tabular}
\end{center}
\label{SE_results}
\end{table}

\begin{table}[t]
    \caption{\small Total voltage deviations and switching required under each scenario.} 
    \begin{center}
    \begin{tabular}{|c|>{\centering\arraybackslash}p{1cm}|>{\centering\arraybackslash}p{1cm}|>{\centering\arraybackslash}p{1.4cm}|>{\centering\arraybackslash}p{1.4cm}|} 
\hline
\multirow{2}{1cm}{Scenario\centering} &\multicolumn{2}{c|}{Number of switching\centering} & \multicolumn{2}{c|}{Total voltage deviation (p.u.)\centering} \\ \cline{2-5}
 &33-bus\centering & 69-bus&33-bus\centering & 69-bus\\
 \hline

i\centering & 2& 2 & 0.84 & 1.31 \\ \hline
ii\centering & 4 & 2 & 0.66 & 1.31\\\hline
iii\centering & 2& 2 & 0.76 & 1.31 \\\hline
iv\centering & 4 & 2 &0.65 & 1.30 \\ \hline
\end{tabular}
\end{center}
\label{V_dev}
\end{table}

\Sajjad{To highlight the contribution of the BO in the reduction of the computational burden of our framework in finding the Stackelberg equilibrium, Table \ref{opt_n} contrasts the optimization counts with and without BO, demonstrating a significant reduction.}

\begin{table}[t]

     \caption{\small Number of optimizations required to compute the Stackelberg equillibrium} 
    \begin{center}
    \begin{tabular}{|c|>{\centering\arraybackslash}p{0.8cm}|>{\centering\arraybackslash}p{0.8cm}|>{\centering\arraybackslash}p{0.8cm}|>{\centering\arraybackslash}p{0.8cm}|} 
\hline
\multirow{2}{1cm}{Scenario\centering} &\multicolumn{2}{c|}{With BO\centering} & \multicolumn{2}{c|}{Without BO\centering} \\ \cline{2-5}
 &33-bus\centering & 69-bus&33-bus\centering & 69-bus\\
 \hline
i, ii, iii\centering & 10& 13 & 32 & 48 \\ \hline
iv \centering & 41 & 69 & 496 & 1128 \\\hline
    \end{tabular}
    \label{opt_n}
    \end{center} \vspace{-2em}
\end{table}

\Sajjad{\subsection{141 Bus Grid}\label{141result}

In this subsection, we extend our simulations to the 141-bus grid to show the scalability of our defensive method to bigger networks. The data for the auxiliary links of this grid is obtained from \cite{helmi2021efficient}. According to this result, when there is an LAA on bus 79, auxiliary line 2-37 is connected, and line 6-37 is disconnected, which maintains the voltage constraints of this grid with a minimum switching.


}
\Sajjad{\subsection{DNs with Inverter-based DERs}

We extend our model by incorporating inverter-based DERs to obtain more realistic models. It is important to note that adding constraints related to DERs changes the nature of the optimization problem: it is no longer an MILP and must instead be formulated and solved as an MISOCP.

The primary effect of incorporating DERs is an increased resilience of the grid against attacks. In other words, the presence of DERs enhances the grid’s ability to withstand more severe attacks. Fig. \ref{33bus_dervsnoder} illustrates the attack configuration on the 33-bus grid with DERs. The red bus represents the targeted node. Notably, when DERs are present in the grid, leaf nodes are no longer necessarily the most attractive targets for the attacker. Instead, the attack strategy becomes closely correlated with the locations of DERs.

To broaden the analysis, Fig. \ref{69bus_dervsnoder} presents the result for the 69-bus grid with DERs. Fig. \ref{vsof33and69} shows the Stackelberg equilibrium voltage profiles for both grids, with and without DERs. In both grids, the integration of DERs brings the voltage profile closer to the reference value of 1p.u.

\begin{figure}[t]
    \centering
    \color{RoyalBlue}      
        \includegraphics[width=0.65\linewidth]{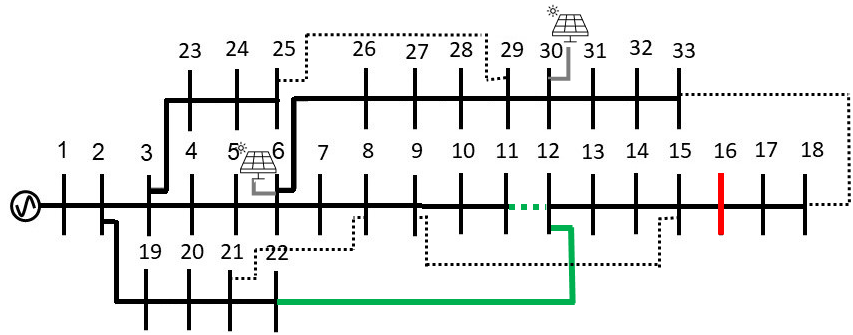}
    \caption{Attack and defense actions in Stackelberg equilibrium for the 33-bus grid with DERs.}
    \label{33bus_dervsnoder}
\end{figure}}

\begin{figure}[t]
\centering
        \includegraphics[width=0.65\linewidth]{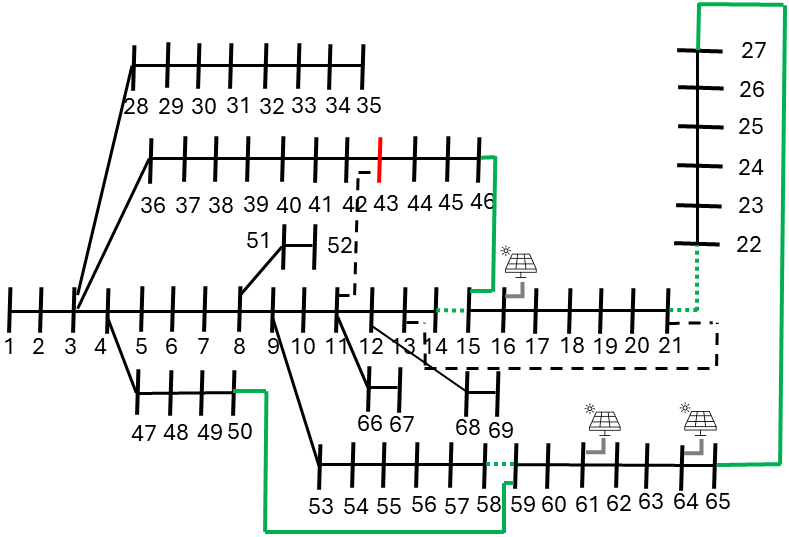}
    \caption{Attack and defense actions in Stackelberg equilibrium for the 69-bus grid with DERs.}
    \label{69bus_dervsnoder}
\end{figure}}

\begin{figure}[t]
    \centering
    \color{RoyalBlue}
   \begin{subfigure}[b]{\linewidth}
        \centering
        \includegraphics[width=0.65\linewidth]{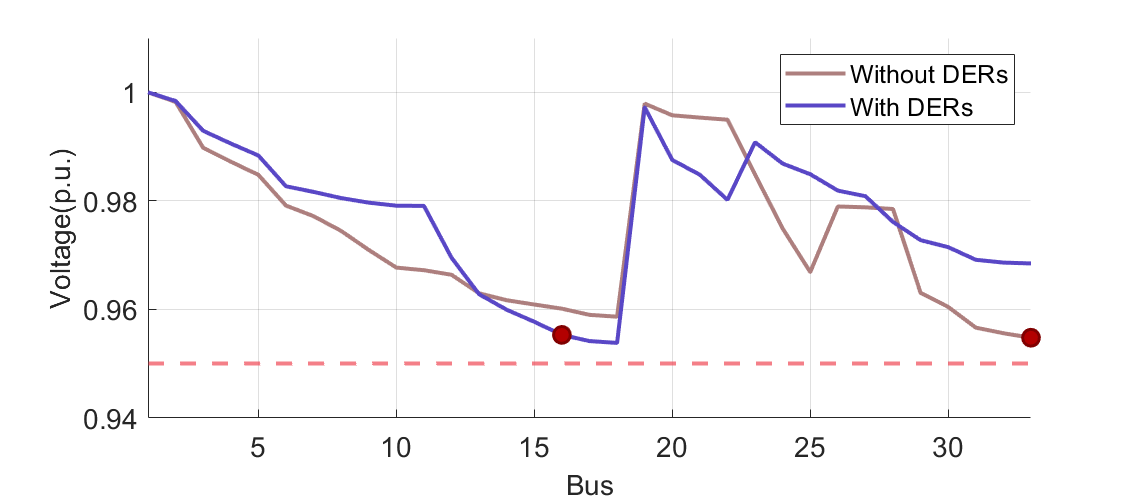}
        \caption{33-bus grid}
        \label{vsof33}
    \end{subfigure}
    \vfill
    \begin{subfigure}[b]{\linewidth}
        \centering
        \includegraphics[width=0.65\linewidth]{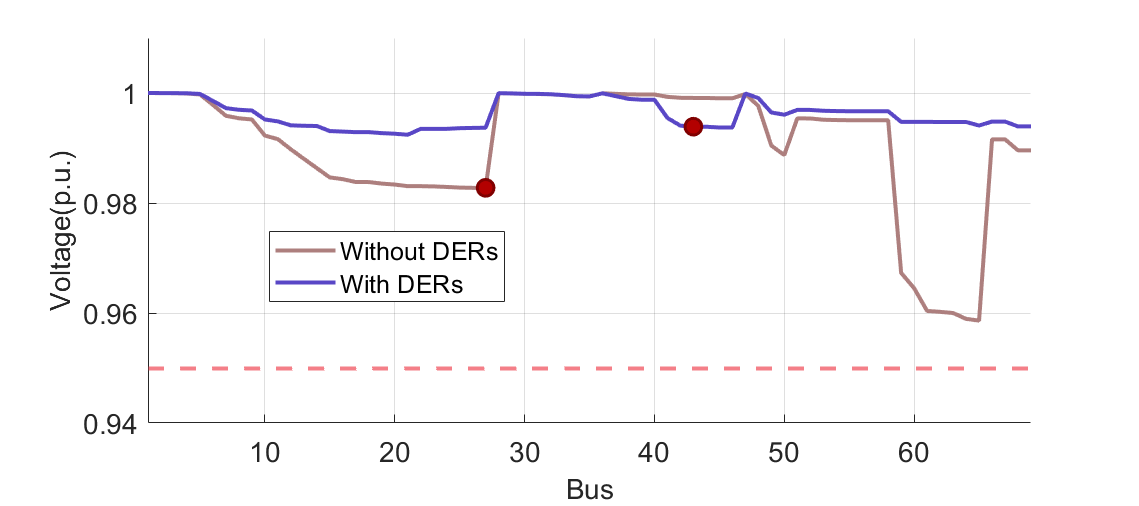}
        \caption{69-bus grid}
        \label{vsof69}
    \end{subfigure}
    \caption{Voltages of grids in Stackelberg equilibrium without and with DERs.}
    \label{vsof33and69}
\end{figure}}

\Sajjad{\subsection{Grid with Less Auxiliary Links}
While the simulation results shown in this section use standard IEEE test grid topologies (including the data on the auxiliary links), we perform additional simulations to show the effectiveness of the proposed defense on the number of auxiliary links. 

For this evaluation, we successively remove three of the four auxiliary links in the 33-bus grid, one at a time. Fig. \ref{voltages-fewerlines} illustrates the voltage profile of the grid when the LAA is conducted at bus 33  when:\\
(a) All four auxiliary links are available;\\
(b) Link 25–29 is unavailable;\\
(c) Links 25–29 and 8–21 are unavailable;\\
(d) Links 25–29, 8–21, and 18–33 are unavailable; and,\\
(e) Links 25–29, 8–21, and 18–33 are unavailable, and the attack magnitude is 20\% bigger than (d).}

\begin{figure}[t]
    \centering
    \includegraphics[width=0.8\linewidth]{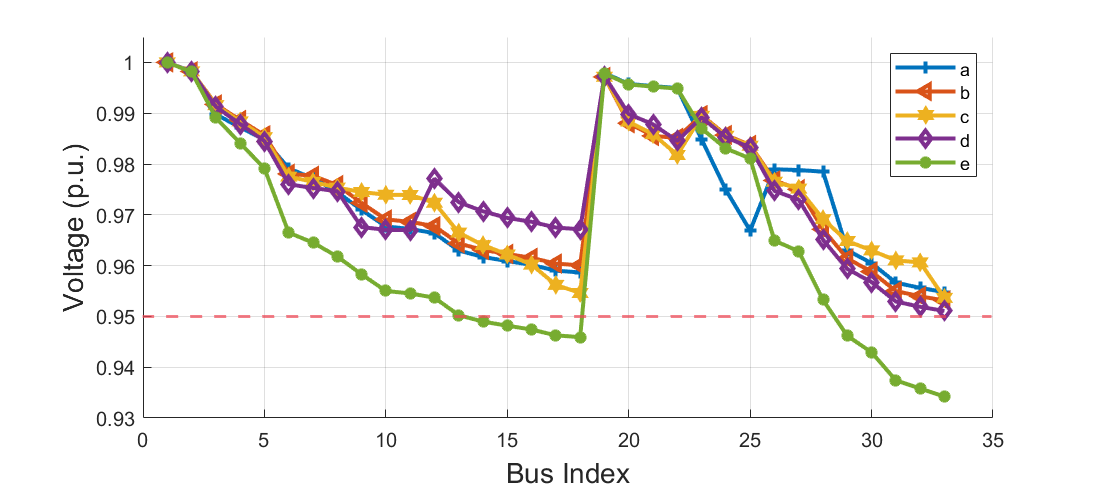}
    \caption{Voltage profile of the 33-bus grid after the LAA on bus 33 and defense when (a) All four auxiliary links are available;
(b) Link 25–29 is unavailable;
(c) Links 25–29 and 8–21 are unavailable;
(d) Links 25–29, 8–21, and 18–33 are unavailable; and (e) Links 25–29, 8–21, and 18–33 are unavailable, and the attack magnitude is 20\% bigger than (d).
}
    \label{voltages-fewerlines}
\end{figure}

\Sajjad{In all scenarios except (e), the reconfiguration-based defense successfully maintains voltage within the specified limits. However, a decrease in the number of auxiliary links results in a slightly worse voltage profile, as the system has fewer reconfiguration options to limit the attack impact. In particular, in scenario (e), with a larger LAA magnitude and with only one auxiliary link, the attack breaches the voltage limits. This shows that while the proposed defense is effective in most practical cases, further research is required on the topic of optimal number and placement of auxiliary links for cyber defense purposes.}

\subsection{Significance of Game-Theoretic Approach} 
We also consider a non-strategic attacker that does not anticipate the defender's actions. 
In this case, first, the attacker launches the attack, which maximizes the total voltage deviation, regardless of potential defensive reactions. Subsequently, the defender solves the optimal reconfiguration problem to mitigate the attack. The results of this approach for the 33-bus test case are presented in Table \ref{non_strategic}. We remark that both scenarios i) and ii) result in the same output. Compared to Table \ref{V_dev} for the strategic attacker, the defender always benefits. Indeed, in scenario i), the total voltage deviation is dropped from $0.84$ p.u. to $0.80$ p.u.; while in ii), the number of switches is reduced from $4$ to $2$. Note that mitigation with less switching is preferred (even with slightly higher voltage deviation). Note that the strategic attacker in the 69-bus grid picks the deepest bus (Bus 27) as the victim, which is similar to the non-strategic attacker's choice.
  \begin{table}[t]
      \caption{\small Non-strategic attacker's preferred actions, obligated switching numbers, and total voltage deviations.} 
      \begin{center}
      \begin{tabular}{|>{\centering\arraybackslash}p{1cm}|c|>{\centering\arraybackslash}p{1.8cm}|>{\centering\arraybackslash}p{1.5cm}|}
  \hline
  \multirow{2}{0.85cm}{Scenario\centering} &\multirow{2}{*}{Attacked bus} & Total voltage deviation (p.u.) & Number of switching \\
   \hline
  i\centering & \centering 18 & 0.80 & 2 \\ \hline
  ii &\centering 18 & 0.80 & 2 \\\hline
  \end{tabular}
  \end{center}
  
  \label{non_strategic}
  \end{table}

\Sajjad{To discuss the efficiency of the approximations introduced in Section \ref{Preliminaries}, we compare the computation time of the implemented MILP model with that of a similar MISOCP model (for grids without DERs). 
All computations were performed using the default Intel(R) Xeon(R) CPU @ 2.20GHz processor on Google Colab. 
 As shown in Table \ref{calculation_time}, a single MISOCP optimization for the 33-bus grid takes 206 seconds, whereas one MILP optimization requires only 19 seconds. This represents a reduction of over 90\% in calculation time, with even more significant reductions observed for larger systems.}

\begin{table}[t]
    \caption{\small \Sajjad{CPU time required to solve different optimizations.}} 
    \begin{center}
    \begin{tabular}{|c|c|c|c|} \hline
        Op. type & 33-bus grid & 69-bus grid & 141-bus grid \\ \hline
         MISOCP& 206 s & 3169 s & $\sim8$ hrs \\ \hline
         MILP & 19 s& 103 s & 1077 s \\ \hline
    \end{tabular}
    \end{center}
    \label{calculation_time} \vspace{-2em}
\end{table}
\Sajjad{\subsection{Real-life Latencies}
Switch actuation time in commercial devices could take up to $50~ms$ \cite{chintNXA}, while communication latencies can add up to $100~ms$  \cite{muyizere2022effects}. In practice, such reconfigurations can be triggered automatically; however, the grid operator may want to ensure that the control actions do not adversely affect the grid. 
To this end, we assume system operators can use a command authentication scheme, which simulates the dynamics of the grid to observe the impact of the control action before its execution in the real system \cite{mashima2018securing}. For an operator using modern-day real-time simulators, this verification can add $50~ms$ \cite{RTDSSimulator2023}. In conclusion, we believe that these actions do not add a significant delay, and the proposed method is still feasible.
}
\section{Conclusions and Future Works}\label{conclusion}
In this paper, we investigate the impact of LAAs on DNs. We derive a set of closed-form expressions for the power flow of DNs to determine bus voltages in the presence of voltage-dependent loads, with or without LAA. Then, we introduce a sequential game-theoretic approach to mitigate LAAs in DNs by network reconfiguration with minimum possible switching \Sajjad{and utilizing the flexibility of inverter-based DERs outputs}. Furthermore, we take into account the uncertainties in the attack localization by introducing a probability distribution over the potentially attacked nodes. To enhance the sustainability and computational speed of the Stackelberg equilibrium, a Bayesian optimization algorithm was implemented to reduce the computational burden. The proposed mitigation scheme is capable of keeping the voltage within the desired constraints. 
Building on our mitigation method, future work will investigate the cyber recovery step, which includes isolating the attack. 
Other interesting extensions include dynamic LAAs and incorporating the transient dynamics of the grid's voltage, among others. \Sajjad{Furthermore, optimal auxiliary lines installation is another potential addition to this work, which should be pursued in the future.}

\Sajjad{Our study provides a foundational one-shot attack-defense model. This work can be extended by future research to develop a multi-stage game-theoretic formulation for analyzing more sophisticated interactions.}

\bibliography{IEEEabrv,Refs} 
\Sajjad{ \appendix \label{appendix}
According to LinDistFlow, the voltage drop between each bus and its parent bus is given by $2(r_{\pi_k,k}p_{\pi_k,k} + x_{\pi_k,k}q_{\pi_k,k})$. Based on graph theory, in a radial network, each bus has a unique path to the root bus. Let $\mathcal{D}_k$ represent the set of buses in the bus $k$'s unique path to the root node. If we sum up all voltage drops along the nodes in $\mathcal{D}_k$, we can conclude:
\begin{equation}
v_k= \sqrt{v_{1}^2 - 2\sum_{i \in \mathcal{D}_k} \left(r_{\pi_i, i}p^{\text{zp}}_{\pi_i, i} + x_{\pi_i, i}q^{\text{zp}}_{\pi_i, i}\right)}.
\end{equation}

Let us denote the victim bus by $ a$. i.e., the demand in bus $a$ is altered due to the LAA. Because of the increased power flow from the root bus to $a$, the voltage drop in the path from bus $1$ to bus $a$ increases accordingly. However, other buses will be impacted as well, and the extent of it for bus $i$ depends on $\mathcal{D}_a \cap \mathcal{D}_i$. The more these two sets have in common, the more the voltage of bus $i$ is affected.

In Fig. \ref{impactfigure}, both green and striped parts are in $\mathcal{D}_a \cap \mathcal{D}_G$. However, only the striped part is in $\mathcal{D}_a \cap \mathcal{D}_Y$. Consequently, when there is an LAA on $a$, bus Y is less impacted than G.
   \begin{figure}
       \centering
       \includegraphics[width=0.4\linewidth]{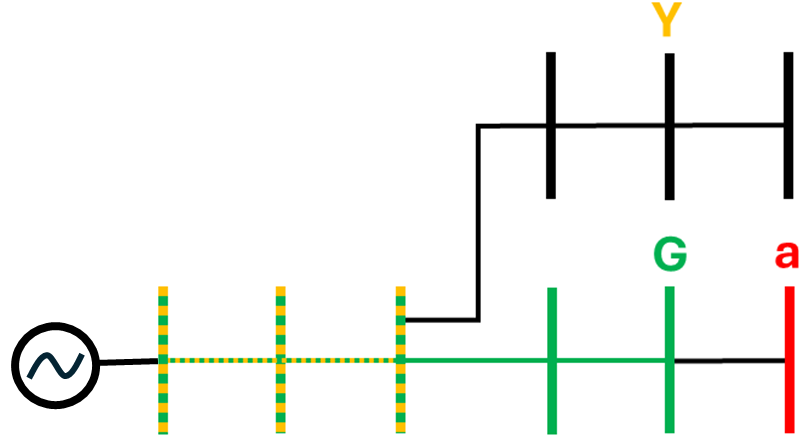}
       \caption{Sample grid} \vspace{-1em}
       \label{impactfigure}
   \end{figure}
   Based on this, the extra voltage drop for bus $k$ is:
   \begin{equation*}
             \sum_{i \in \mathcal{D}_a \cap \mathcal{D}_k} 2p_a^A r_{\pi_i},i + 2q_a^A x_{\pi_i,i}.
   \end{equation*}
   To have a more compact equation, we introduce $r_{k,a} = \sum_{i \in \{\mathcal{D}_a \cap\mathcal{D}_k\}} r_{\pi_i, i},$ and $ x_{k,a} = \sum_{i \in \{\mathcal{D}_a \cap\mathcal{D}_k\}} x_{\pi_i, i}$. As a result, the final equation for the nodal voltage while LAA is:
   \begin{equation}
     \sqrt{v_{1}^2 - 2\sum_{i \in \mathcal{D}_k} \left(r_{\pi_i, i}p^{zp}_{\pi_i, i} + x_{\pi_i, i}q^{zp}_{\pi_i, i}\right) - 2p^{A}_ar_{k,a} - 2q^{A}_ax_{k,a}}  
   \end{equation}}

 \vskip -2\baselineskip plus -1fil
 \begin{IEEEbiographynophoto}
     {Sajjad Maleki}~(Graduate Student Member, IEEE) received his  BSc. degree in electrical power engineering in 2017 and his MSc. in 2020 in power systems engineering, both from the University of Tabriz, Iran. He is currently a PhD student jointly at the University of Warwick, UK and CY Cergy Paris University, France. His research interests include cybersecurity, optimization and game theory applications in power grids. His work was selected as the best paper at the IEEE SmartGridComm 2024 conference.
 \end{IEEEbiographynophoto}
 \vskip -2\baselineskip plus -1fil
 \begin{IEEEbiographynophoto}
 {E. Veronica Belmega}~(IEEE S’08, M’10, SM’20) is
 a full professor at the Universite Gustave Eiffel and LIGM laboratory, Marne-la-Vall\'ee, France, since May 2022. Previously, she was an associate professor (MCF HDR) with ENSEA graduate school (Sep.
 2011 - Apr. 2022) and Deputy Director of ETIS
 laboratory (Jan. 2020 - Apr. 2022), Cergy, France.
 She received the M.Sc. (engineering) degree from
 the University Politehnica of Bucharest, Romania,
 in 2007, and the M.Sc. and Ph.D. degrees both from
 the University Paris-Sud 11, Orsay, France, in 2007
 and 2010. From 2010 to 2011, she was a post-doctoral researcher at Princeton University and Supelec. In 2015-2017, she was a visiting researcher at Inria, France. In 2009, she received the
 French L’Or\'eal – UNESCO – French Academy of Science fellowship and, in 2021, she received the CY Alliance award For Women in Science, France. She is the co-recipient of the Best Paper Awards at ICL-GNSS 2023 and IEEE SmartGridComm 2024. She currently serves as Area Editor for the IEEE TRANSACTIONS ON MACHINE LEARNING
 IN COMMUNICATIONS AND NETWORKING. 
 Her research interests lie in convex
 optimization, game theory and machine learning applied to distributed wireless
 communication and power networks.
 \end{IEEEbiographynophoto}
 \vskip -2\baselineskip plus -1fil
 \begin{IEEEbiographynophoto}{Charalambos Konstantinou}~(S'11-M'18-SM'20) is an Associate Professor of Electrical and Computer Engineering with the Computer, Electrical and Mathematical Science and Engineering Division (CEMSE), King Abdullah University of Science and Technology (KAUST), Thuwal, Saudi Arabia. He received the M.Eng. degree in ECE from the National Technical University of Athens (NTUA), Greece, and the Ph.D. degree in Electrical Engineering from New York University (NYU), NY, USA. His research interests include critical infrastructures security and resilience with special focus on smart grid technologies, renewable energy integration, and real-time simulation. He is currently serving as Associate Editor of IEEE Transactions on Smart Grid (TSG) and IEEE Transactions on Industrial Informatics (TII).
 \end{IEEEbiographynophoto}
 \vskip -2\baselineskip plus -1fil
 \begin{IEEEbiographynophoto}{Subhash Lakshminarayana} (S'07, M'12, SM'20) is an assistant professor at the School of Engineering, University of Warwick, UK. Previously, he worked as a researcher at the Advanced Digital Sciences Center (ADSC) in Singapore between 2015-2018, a joint post-doctoral researcher at Princeton University and the Singapore University of Technology and Design (SUTD) between 2013-2015. He received his Ph.D. from the Alcatel Lucent Chair on Flexible Radio and the Department of Telecommunications at École supérieure d'électricité, France in 2013, M.S. degree in Electrical and Computer Engineering from The Ohio State University in 2009 and B.S. from Bangalore University, India. 
 His research interests include cyber-physical system security (power grids and urban transportation) and wireless communications. His works have been selected among the Best conference papers on integration of renewable \& intermittent resources at the IEEE PESGM - 2015 conference, and the ``Best 50 papers" of IEEE Globecom 2014 conference. 
 \end{IEEEbiographynophoto}
\end{document}